# The Chemistry of Spinel Ferrite Nanoparticle Nucleation, Crystallization, and Growth

Henrik L. Andersen,* Cecilia Granados-Miralles, Kirsten M. Ø. Jensen, Matilde Saura-Múzquiz, and Mogens Christensen

**ABSTRACT:** The nucleation, crystallization, and growth mechanisms of $MnFe_2O_4$, $CoFe_2O_4$, $NiFe_2O_4$, and $ZnFe_2O_4$ nanocrystallites prepared from coprecipitated transition metal (TM) hydroxide precursors treated at sub-, near-, and supercritical hydrothermal conditions have been studied by *in situ* X-ray total scattering (TS) with pair distribution function (PDF) analysis, and *in situ* synchrotron powder X-ray diffraction (PXRD) with Rietveld analysis. The *in situ* TS experiments were carried out on 0.6 M TM hydroxide precursors prepared from aqueous metal chloride solutions using 24.5% $NH_4OH$ as the precipitating base. The PDF analysis reveals equivalent nucleation processes for the four spinel ferrite compounds under the studied hydrothermal conditions, where the TMs form edge-sharing octahedrally coordinated hydroxide units (monomers/dimers and in some cases trimers) in the aqueous precursor, which upon hydrothermal treatment nucleate through linking by tetrahedrally coordinated TMs. The *in situ* PXRD experiments were carried out on 1.2 M TM hydroxide precursors prepared from aqueous metal nitrate solutions using 16 M NaOH as the precipitating base. The crystallization and growth of the nanocrystallites were found to progress via different processes depending on the specific TMs and synthesis temperatures. The PXRD data show that $MnFe_2O_4$ and $CoFe_2O_4$ nanocrystallites rapidly grow (typically <1 min) to equilibrium sizes of 20−25 nm and 10−12 nm, respectively, regardless of applied temperature in the 170−420 °C range, indicating limited possibility of targeted size control. However, varying the reaction time (0−30 min) and temperature (150−400 °C) allows different sizes to be obtained for $NiFe_2O_4$ (3−30 nm) and $ZnFe_2O_4$ (3−12 nm) nanocrystallites. The mechanisms controlling the crystallization and growth (nucleation, growth by diffusion, Ostwald ripening, etc.) were examined by qualitative analysis of the evolution in refined scale factor (proportional to extent of crystallization) and mean crystallite volume (proportional to extent of growth). Interestingly, lower kinetic barriers are observed for the formation of the mixed spinels ($MnFe_2O_4$ and $CoFe_2O_4$) compared to the inverse ($NiFe_2O_4$) and normal ($ZnFe_2O_4$) spinel structured compounds, suggesting that the energy barrier for formation may be lowered when the TMs have no site preference.

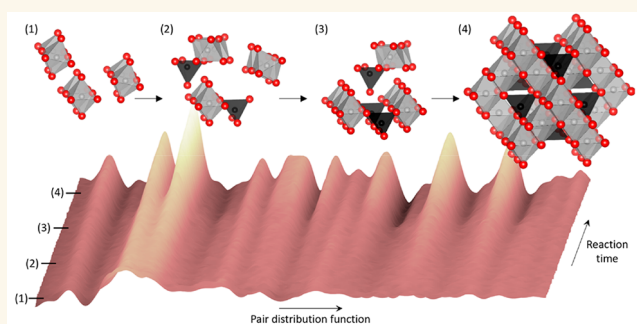

## INTRODUCTION

In recent years, spinel ferrite ($MFe_2O_4$, M = Mn, Fe, Co, Ni, Zn, etc.) nanoparticles have been receiving increasing research interest due to current and future applications in magnetically recoverable nanocatalysts,[1,2] MRI contrast agents,[3,4] hyperthermia cancer treatment,[5,6] magnetic exchange-spring nanocomposites,[7,8] neuromorphic spintronics,[9,10] drug delivery,[11,12] and many more applications.[13] Here, the magnetic spinel-structured ferrite compounds benefit from their low cost, excellent resistance to corrosion, tunable properties, and good magnetic performance.[14,15] The spinel ferrite compounds crystallize in the spinel structure (space group $Fd\bar{3}m$), illustrated in Figure 1. The different divalent cations, $M^{2+}$, are known to exhibit different affinities for the specific crystallographic sites resulting in formation of either normal spinel structures (all $M^{2+}$ occupying all tetrahedral 8a Wyckoff sites), inverse spinel structures (all $M^{2+}$ occupying half the octahedral 16d Wyckoff sites), or mixed spinels with a fraction, $x$, of the $Fe^{3+}$ ions (called the inversion degree) occupying the tetrahedral sites, $[M^{2+}_{1-x}Fe^{3+}_x]^{tet}[M_xFe^{3+}_{2-x}]^{oct}O_4$. For larger/bulk crystallites, the thermodynamically stable cation distribution is normal ($x = 0$) for $ZnFe_2O_4$, mixed for $MnFe_2O_4$, and inverse ($x = 1$) for $CoFe_2O_4$ and $NiFe_2O_4$,[16] but nanosized crystallites have been reported to exhibit a variety of inversion degrees.[17−24] In their thermodynamically stable bulk forms and at room temperature, $MnFe_2O_4$ and $NiFe_2O_4$ are soft ferrimagnets, $CoFe_2O_4$ is a hard ferrimagnet, and $ZnFe_2O_4$ is paramagnetic, however, ultrafine nanocrystals of the compounds will exhibit superparamagnetic behavior below their blocking temperature. Thus, the magnetic properties and performance are governed both by the composition, cation distribution (inversion degree), and nanoparticle sizes, providing several flexible handles to tune the materials performance.

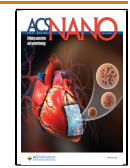

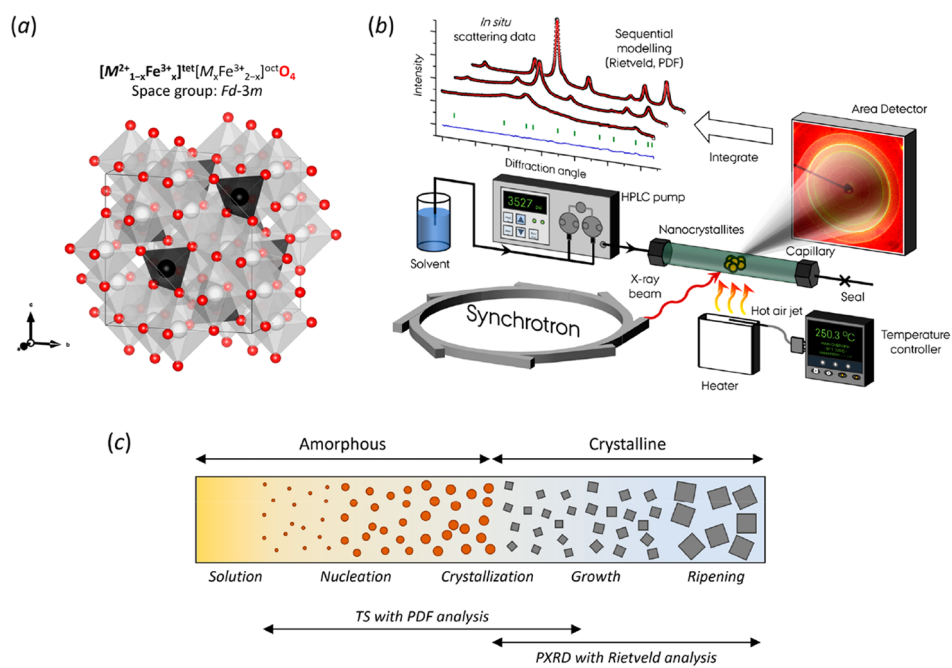

Figure 1. (a) Illustration of the cubic spinel structure ($Fd\bar{3}m$) with oxide ions in red, tetrahedrally coordinated cations in black, and octahedrally coordinated cations in white. Illustration made with VESTA.[25] (b) Schematic illustration of the experimental setup used for the *in situ* powder X-ray diffraction (PXRD) and total scattering (TS) studies of hydrothermal nanoparticle formation and growth. (c) Illustration of the different stages of hydrothermal nanoparticle formation, crystallization, and growth and the corresponding application ranges of the *in situ* scattering methods used for the characterization.

To improve the performance of spinel ferrite-based materials, it is crucial to have control over both the crystal- and microstructural characteristics during synthesis. In addition, for the material to have commercial/industrial relevance, the employed synthesis method must simultaneously be simple, cheap, scalable, environmentally friendly, and energy efficient. In this context, the solvo- and hydrothermal methods are excellent approaches, which fulfill the aforementioned requirements. Numerous studies of hydrothermal nanoparticle synthesis have demonstrated how product characteristics can be tuned by simple adjustments to the applied reaction conditions, such as temperature, heating rate, precursor concentration, pH, pressure, and reaction time, and the method is already widely used for industrial-scale production of a large variety of functional materials.[26−30] Consequently, understanding the inorganic chemistry of solvothermal nanoparticle formation and growth is of great importance for numerous areas of science and technology. In this context, *in situ* characterization methods, which are becoming increasingly available at large radiation facilities, have revealed detailed mechanistic insight into the formation and kinetics of functional materials synthesis in a diverse range of fields, such as Li-ion battery cathodes,[31−33] photocatalysts,[34,35] nonlinear optical materials,[36] multifunctional oxides,[37,38] $TiO_2$ nanocatalysts,[39] high-entropy alloys,[40] materials for artificial bone or teeth,[41] nanostructured magnets,[42,43] as well as geologically and environmentally important compounds.[44,45]

Traditionally, the nucleation and growth processes involved in solvothermal nanoparticle formation have been discussed principally based on thermodynamic considerations on a particle level, ignoring the atomic-scale chemical nature of the systems.[46] However, over the past decade, experiments utilizing total scattering (TS) with pair distribution function (PDF) to study the solvothermal nucleation mechanisms of a number of nanostructured intermetallic and oxide systems have made it increasingly clear that the early stage precrystalline nucleation of nanoparticles under solvothermal conditions often involve more complex mechanisms.[46] In general, the studies have revealed mechanisms involving various precrystalline clustering steps and reconfigurations of local structure prior to crystallization. In 2012, the first *in situ* study of solvothermal nanoparticle formation using combined analysis of TS and powder X-ray diffraction (PXRD) data examined the hydrothermal formation of $SnO_2$ from aqueous $SnCl_4$ precursor solutions.[37] The study showed how *mer*-$[SnCl_3(H_2O)_3]^+$ precursor complexes gradually disproportionate into $[Sn(H_2O)_6]^{4+}$ units, which cluster together to form the rutile structured $SnO_2$ nanoparticle product.[37]

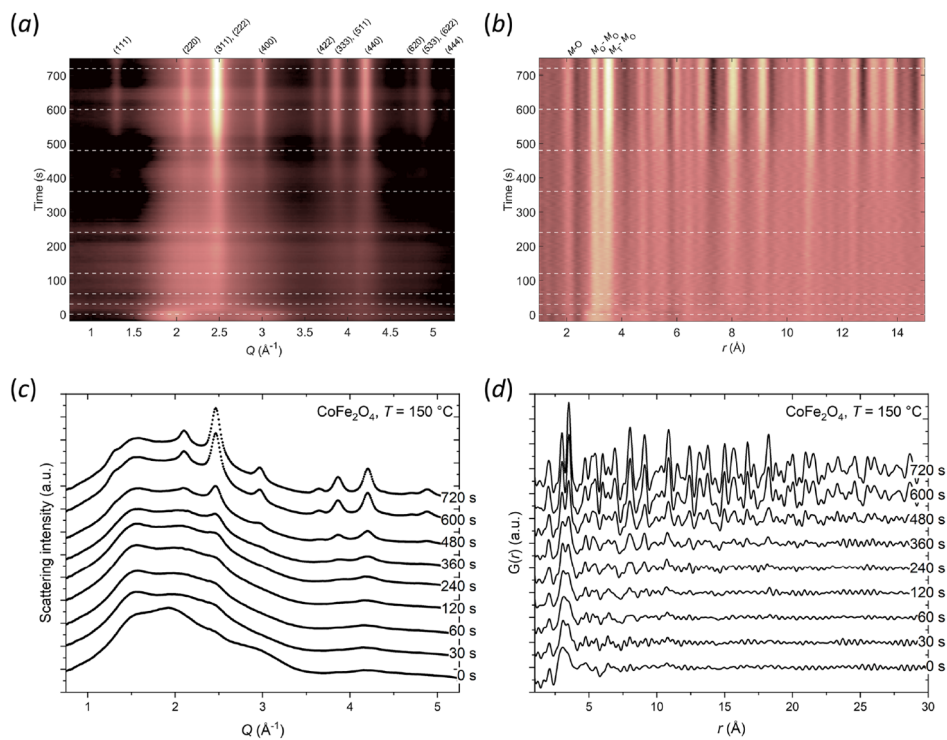

Figure 2. (a) Background-subtracted and normalized contour plot of the low-Q region of the time-resolved total scattering data obtained during formation of CoFe$_2$O$_4$ nanoparticles by hydrothermal treatment of the NH$_4$OH precipitated precursor at 150 °C and 110 bar. The hkl reflections from the spinel structure have been labeled above the contour plot. (b) Corresponding contour plot of the PDFs. (c) Low-Q regions of selected in situ diffraction data sets (background not subtracted) obtained after the indicated hydrothermal treatment times. (d) Corresponding in situ PDFs.

Another example is the study of the hydrothermal formation of CeO$_2$ nanoparticles from a Ce(NH$_4$)$_2$(NO$_3$)$_6$ aqueous solution, which showed the presence of a larger dimeric precursor complex, which upon heating is converted into CeO$_2$ likely by clustering followed by fast internal restructuring prior to growth.[47] Additional examples of nucleation mechanisms of solvothermal nanoparticle systems that have been studied include different oxide nanocrystallites, such as HfO$_2$,[48] Nb$_2$O$_5$,[49] WO$_3$,[50] TiO$_2$,[51] ZrO$_2$,[52,53] ZnWO$_4$,[54] as well as metal halide nanoparticles, e.g., Ir$_x$Cl$_y$,[55] metallic nanoparticles, e.g., Pt/Pt$_3$Gd,[56] Pd−Pt,[40,57] PdIn,[58] FeSb$_2$/FeSb$_3$,[59] and high entropy alloy nanoparticles.[60] In several of the mentioned metal oxide systems, the precursors were observed to contain dimeric or trimeric metal oxide/hydroxide polyhedra, which "polymerize" and/or reconfigure to form larger clusters during nucleation.

In the present study, the nucleation and crystallization of magnetic spinel ferrite nanoparticles (MFe$_2$O$_4$, M = Mn, Co, Ni, Zn) from different coprecipitated transition metal (TM) hydroxide precursors treated at sub-, near-, and supercritical hydrothermal conditions have been studied by in situ synchrotron X-ray TS with PDF analysis and in situ synchrotron PXRD with Rietveld analysis, respectively, using a custom built in situ setup (see Figure 1b).[61−63] Studying the chemical reactions in situ reduces the time needed to cover parameters space (specifically the reaction time component) and allows observation of the fundamental atomic-scale chemistry that takes place between inorganic species during nucleation, crystallization and growth of the nanoparticles,[46,64,65] an aspect that is still relatively poorly understood compared to reactions in organic chemistry.[45,66] As illustrated in Figure 1c, in situ TS with PDF analysis is ideal for examining the local structure of the amorphous precursor clusters and disordered nanoparticles in the early stages of the hydrothermal nanoparticle formation,[67] while in situ PXRD with Rietveld and peak profile analysis provides a more robust view of the evolution in average long-range atomic structure and crystallite sizes. Here, in situ TS experiments were carried out on ∼0.6 M TM hydroxide precursors prepared from aqueous metal chloride solutions using 24.5% NH$_4$OH as the precipitating base. Using PDF analysis, we elucidate the precrystalline nucleation mechanisms, which are found to involve the formation of edge-sharing octahedrally coordinated transition metal (TM) hydroxide units in the aqueous precursor that subsequently nucleate through linking by tetrahedrally coordinated TMs. The in situ PXRD experiments were carried out on ∼1.2 M TM hydroxide precursors prepared from aqueous metal nitrate solutions using 16 M NaOH as the precipitating base. Using Rietveld and peak profile analysis, we examine the crystallization of the spinel ferrite nanoparticles and demonstrate how their sizes can be controlled by varying the spinel ferrite composition (type of divalent cation, M) and/or through modifications to the precursor composition/preparation route and hydrothermal synthesis temperature.

## RESULTS

In the present study, two different routes have been employed for the coprecipitation of the metal hydroxide precursors used for the in situ PXRD and TS experiments (details can be found in the Methods section). A weaker 24.5% NH$_4$OH base was used for the *in situ* TS experiment precursor preparation (NH$_4$OH route), while a stronger 16 M NaOH base was used for the PXRD experiments precursor (NaOH route). Using the NaOH route, i.e., the stronger base, allows a precursor pH of 14 to be reached with relatively limited dilution of the precursor mixture. This makes it ideal for *in situ* scattering experiments as it allows higher metal ion concentration (~1.2 M) to be present in the illuminated volume, thereby yielding a better scattering signal-to-noise ratio. However, for the *in situ* total scattering experiments it was necessary to use amorphous fused silica capillaries (rather than the chemically and mechanically resilient single-crystal sapphire capillaries used for the *in situ* synchrotron PXRD experiments) to allow deconvolution of the sample and background scattering contributions. Fused silica capillaries tolerate lower pressures and are degraded by harsh bases such as concentrated NaOH,[61] and consequently, the method using chemically weaker NH$_4$OH base to induce the TM hydroxide precipitation in the precursor preparation (metal ion concentration of ~0.6 M), as well as lower pressure (110 rather than 250 bar), were employed for the *in situ* total scattering experiments. Therefore, given the differences in synthesis parameters and conditions, the observations from the PXRD experiments do not necessarily represent the true continuation of the nucleation mechanisms and early crystallization paths observed in the TS data. This is detailed further in the Discussion section.

**Nucleation Mechanism—*In Situ* Total Scattering and PDF Analysis.** In TS and PDF data analysis (as opposed to conventional diffraction data analysis), the scattering signals arising from both the short- and long-range order in the sample (i.e., diffuse and Bragg scattering, respectively) are analyzed thereby allowing extraction of structural information from both amorphous, nanosized and crystalline structures.[67] Thus, to study the precrystalline nucleation mechanism of the four types of $M$Fe$_2$O$_4$ nanoparticles under hydrothermal conditions, we conducted *in situ* total scattering experiments with the precursors prepared via the NH$_4$OH route described in the Methods section. The precursor was loaded into the *in situ* capillary reactor setup illustrated in Figure 1b (see description in Methods), and TS experiments were conducted with reactor temperatures ranging between 150 and 250 °C at a pressure of 110 bar. Representative time-resolved $Q$-space scattering data, and the corresponding evolution in atomic PDFs during the hydrothermal formation of CoFe$_2$O$_4$ nanoparticles at 150 °C are shown in Figure 2a and b, respectively. The observed changes in the total scattering and PDF data can be correlated to changes in the atomic structure of the species in the precursor solution and the formed product. Initially, before the heating is initiated at $t = 0$ s, the Co and Fe metal-hydroxide suspension only gives rise to broad diffuse $Q$-space scattering peaks that are difficult to discern from the background signal indicating a largely amorphous precursor structure (see Figure 2c). However, in the corresponding PDF (see Figure 2b and d), a few vague features can be seen at low $r$, i.e., up to approximately 4 Å, indicating the presence of very local order. This includes a peak at 2 Å, which agrees with typical Fe−O or Co−O bonds, as well as broader features between 2.5 and 3.5 Å. The PDF peak intensities are determined by the relative concentration of the associated atomic pair separated by the distance, $r$, and the scattering power of the involved atoms. Consequently, changes in relative PDF peak intensities and positions are related to structural changes, while peaks appearing at higher $r$ indicate growth of the coherently scattering domains.

As heating is started, an immediate change is observed in the scattering signal with the diffuse scattering features at ~2 Å$^{-1}$ and 3 Å$^{-1}$ disappearing and a new broad peak becoming visible at ~2.5 Å$^{-1}$, which corresponds to the position of the main (311) reflection of the spinel ferrite structure. Similarly, in the PDF data, the Fe/Co−O peak at 2 Å grows/sharpens and the features at higher $r$ sharpen with distinct peaks at ~3 and 3.5 Å appearing, which correspond to the nearest neighbor distances between metal ions in edge-sharing octahedra ($M_O$−$M_O$) and corner-sharing tetrahedra and octahedra ($M_T$−$M_O$) in the spinel structure, respectively.

Following this initial nucleation step, for the next ~300 s only minor changes and fluctuations in scattering intensity are observed in the total scattering data likely due to sample movement, temperature fluctuations and/or variations in beam intensity. It is not until after ~360 s that a seemingly second step in the nucleation/crystallization mechanism commences. At this point, additional diffraction peaks from the spinel structure become clearly visible in the $Q$-space data. Similarly, in the PDF data, no obvious changes are observed for the initial ~240 s. However, after 240 s, a gradual increase in the structural coherence length is observed with new peaks arising at higher $r$ along with a sharpening of the individual peaks in the PDF, and after 600 s the structural order clearly extends beyond 30 Å.

The atomic pairs in the bulk CoFe$_2$O$_4$ spinel structure contributing to the PDF peaks between 0−4 Å are listed in Table 1, and Figure 3a shows the low-$r$ region (<11 Å) for selected PDF data frames collected at 150 °C along with vertical lines indicating characteristic interatomic correlations. For the precursor PDF ($t = 0$ s), only very local correlations (below 4 Å) are observed, with the main features in the PDF corresponding to the nearest neighbor Fe/Co−O bonds (~2 Å), the $M_O$−$M_O$ pair (~3 Å), and potentially a shoulder associated with the $M_T$−$M_O$ (~3.5 Å) pair, which are all consistent with equivalent distances in the final spinel CoFe$_2$O$_4$ structure. In addition, a contribution is observed at ~2.8 Å (see blue arrow in Figure 3a), which is visible as a shoulder on the characteristic edge-sharing octahedral-octahedral transition metal coordination (~3 Å) peak. Attempts at identifying the origin of this 2.8 Å precursor peak were unsuccessful. While it does correspond to a characteristic O–O distance in the CoFe$_2$O$_4$ spinel structure (2.82 Å), this correlation would not be expected to yield as high an intensity relative to the other peaks. Instead, it is likely related to the local structure of iron(III) oxyhydroxide, FeOOH, and/or cobalt hydroxide, Co(OH)$_2$, clusters in the precursor (see discussion below).

**Table 1. List of Low-$r$ Atomic Correlations Giving Rise to Peaks in the PDF[a]**

| atom pair | distance (Å) | multiplicity |
|---|---|---|
| $M_T$−O | 1.91 | 32 |
| $M_O$−O | 2.05 | 96 |
| O−O | 2.82 | 48 |
| $M_O$−$M_O$ | 2.97 | 48 |
| O−O | 2.97 | 96 |
| O−O | 3.11 | 48 |
| $M_T$−$M_O$ | 3.48 | 96 |
| $M_T$−O | 3.50 | 96 |
| $M_O$−O | 3.54 | 32 |
| $M_T$−$M_T$ | 3.63 | 16 |
| $M_O$−O | 3.66 | 96 |

[a]Values have been calculated from the previously reported spinel structure of CoFe$_2$O$_4$ nanocrystallites.[68]

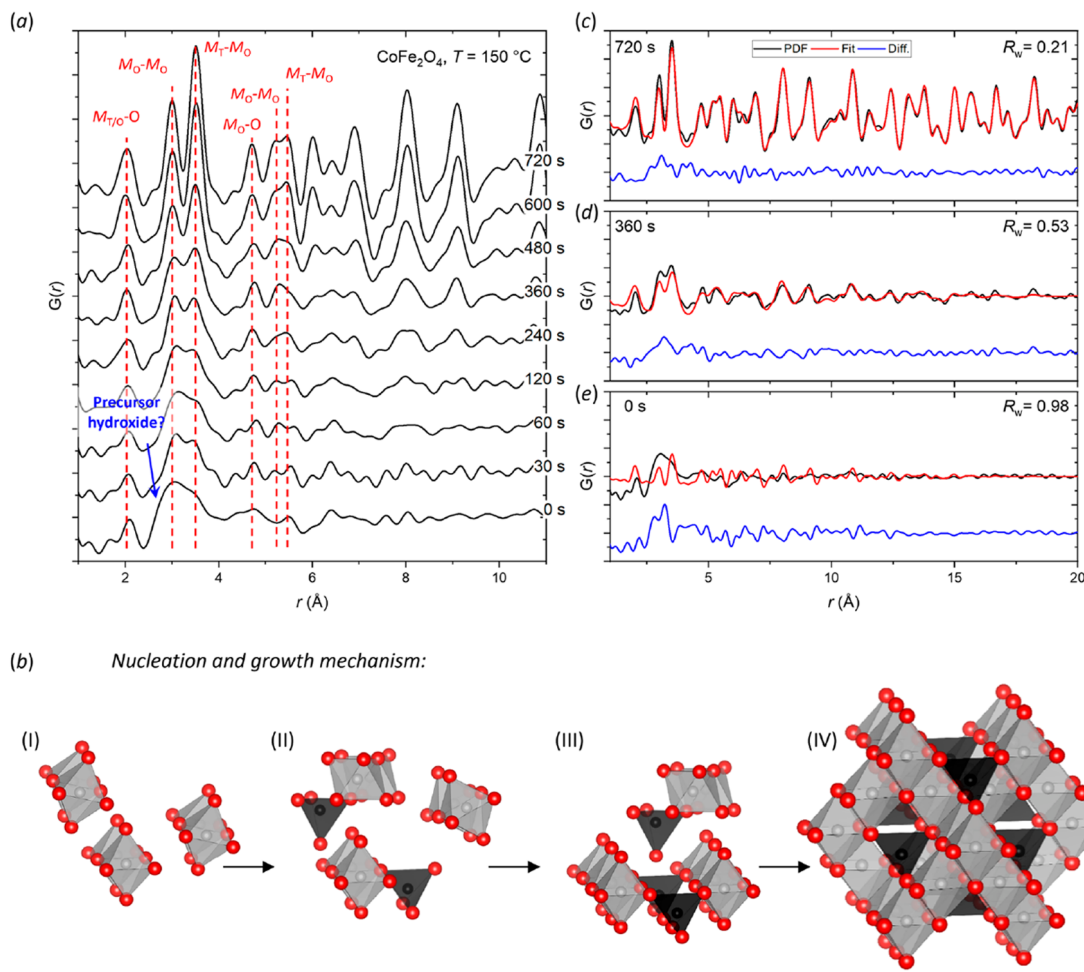

Figure 3. (a) Stacked plot of the low-$r$ region of selected PDF frames collected during CoFe$_2$O$_4$ nanoparticle synthesis at 150 °C. The red lines mark the main characteristic interatomic distances between oxygen (O), tetrahedrally coordinated Co or Fe ($M_T$), and octahedrally coordinated Co or Fe ($M_O$) in the CoFe$_2$O$_4$ spinel structure. The blue arrow highlights the peak/shoulder arising from a precursor interatomic distance at approximately 2.8 Å. (b) Illustration of proposed mechanism for nucleation and growth based on the observed evolution in the PDF data (see description in text). The red spheres indicate oxygen atoms while the white and black spheres indicate octahedrally and tetrahedrally coordinated metal atoms, respectively. (c–e) Representative PDF fits in the range 1–20 Å to data obtained after (c) 720 s, (d) 360 s, and (e) 0 s.

As heating is commenced, the ∼2.8 Å precursor shoulder disappears as the metal hydroxide species are consumed to form CoFe$_2$O$_4$ primary nuclei. With time, the Fe/Co–O peak at ∼2 Å grows/sharpens and the features at higher $r$ sharpen with distinct peaks at ∼3 Å and ∼3.5 Å appearing, which correspond to the $M_O$–$M_O$ and $M_T$–$M_O$ pairs, respectively. Simultaneously, peaks corresponding to increasingly longer-range atomic correlations in the spinel structure appear in the PDF indicating cluster growth. It can be observed how the relative intensities of the $M_O$–$M_O$ (∼3 Å) and $M_T$–$M_O$ (∼3.5 Å) peaks clearly change during the nucleation and growth of the particles. At $t = 0$ s the $M_T$–$M_O$ peak is relatively weaker than the $M_O$–$M_O$ peak (i.e., the $M_T$–$M_O$:$M_O$–$M_O$ intensity ratio approximately 4:5), but it gradually grows to become more intense as the longer-range spinel structure is formed. Finally, after 720 s of hydrothermal treatment at 150 °C, the relative intensities of the two peaks have reversed with the relative ratio being approximately 4:3. Thus, the data implies a nucleation mechanism, where edge-sharing octahedral hydroxide dimer clusters, which are present already prior to heating, are progressively linked through a condensation reaction (i.e., through elimination of H$_2$O from the hydroxides) by tetrahedrally coordinated metal ions as the particles grow. This is equivalent to our previously observed nucleation mechanism for spinel iron oxide particles prepared by hydrothermal treatment of aqueous ammonium iron(III) citrate solutions.[69] The similar behavior could indicate a general nucleation mechanism for spinel structured ferric nanoparticles, irrespective of whether they are prepared from salt solutions or coprecipitated hydroxides.

In summary, the PDF data indicate that upon applying heat, the CoFe$_2$O$_4$ nanoparticle nucleation likely occurs by the following mechanism and as illustrated in Figure 3b. (**I**) Initially, the precursor consists of octahedrally coordinated Fe/Co hydroxide dimers and potentially to a lesser extent of monomers of tetrahedrally coordinated Fe/Co. (II) Upon heating, the clusters are linked by tetrahedrally coordinated metal ions via a condensation reaction. (III) The particles continue to grow by further incorporation of the tetrahedrally coordinated Fe/Co, as evident from the increase in the ratio between the $M_T$–$M_O$ (3.5 Å) and $M_O$–$M_O$ (3 Å) PDF peaks and the gradual appearance of longer-range correlations. (IV) As heating continues, the clusters link and crystallize to form larger crystallites with the final long-range crystalline spinel structure.

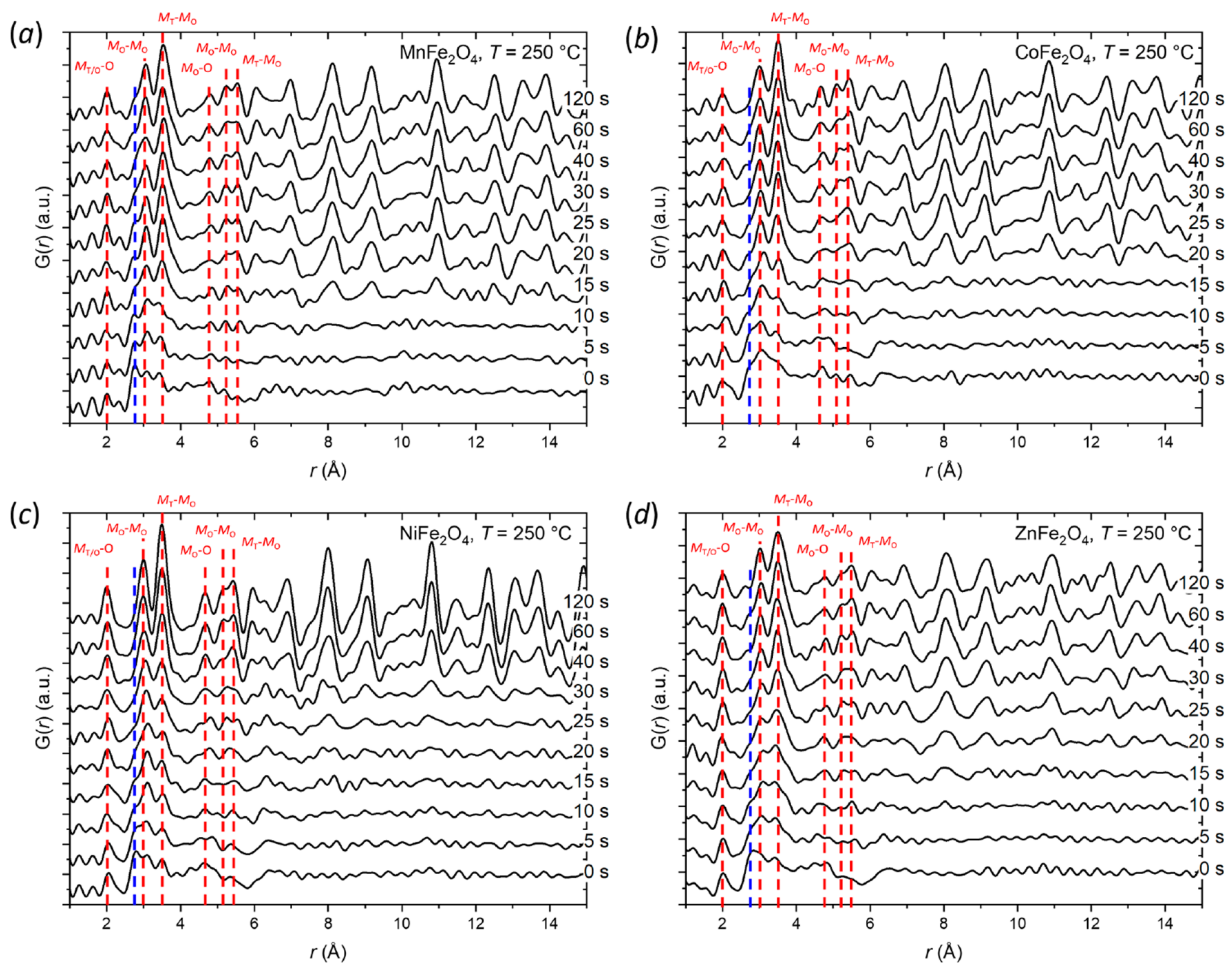

Figure 4. Selected PDF data sets collected within the initial 120 s of nanoparticle synthesis at 250 °C for (a) $MnFe_2O_4$, (b) $CoFe_2O_4$, (c) $NiFe_2O_4$, and (d) $ZnFe_2O_4$. The red lines indicate some characteristic interatomic distances between oxygen (O), tetrahedrally coordinated TMs ($M_T$), and octahedrally coordinated TMs ($M_O$) in the spinel structure. The blue line indicates an unknown precursor peak/shoulder (~2.8 Å), which is observed in all samples at $t = 0$ s.

Figure 3c–e show examples of real-space Rietveld fits based on the $CoFe_2O_4$ spinel structure in space group $Fd\bar{3}m$ to the PDF data obtained at different times, i.e., after 0, 360, and 720 s, during the hydrothermal synthesis. The data obtained after both 360 and 720 s are clearly consistent with the spinel structure but with different particle diameters of 21(6) Å and 43(10) Å, respectively, according to the refinement. However, the observed local structure of the precursor (0 s), which only extends a few Å, could not be satisfactorily refined by the model based on the spinel structure. Numerous combinations of possible precrystalline Fe and Co cluster structures can give rise to the PDF data observed for the precursor as well as during the early stage nucleation process (<120 s). Consequently, despite the clear evidence of local structure

being present, no unambiguous determination of the exact precrystallization cluster structure can be deduced from the present data.

Figure 4a–d show selected data frames during the first 120 s of nanoparticle synthesis for the four studied spinel ferrite systems, i.e., $MnFe_2O_4$, $CoFe_2O_4$, $NiFe_2O_4$, and $ZnFe_2O_4$, under hydrothermal conditions at 110 bar and 250 °C. Generally, very similar observations (i.e., evidence of equivalent nucleation mechanisms to the one described above for $CoFe_2O_4$ at 150 °C) are observed for all four studied ferrite nanoparticle syntheses, with the nucleation reaction progressing faster at this higher synthesis temperature. The unknown ~2.8 Å precursor peak ($t = 0$ s) is present for all four compositions (see dashed blue lines Figure 4a–d). Notably, the signal appears relatively higher in the $MnFe_2O_4$ and $NiFe_2O_4$ precursors compared the $CoFe_2O_4$ and $ZnFe_2O_4$ precursors. This could be indicative of some primary nuclei already having formed in the $CoFe_2O_4$ and $ZnFe_2O_4$ precursors, potentially helped by the exothermic reaction occurring when adding the $NH_4OH$ to the salt solutions in the precursor preparation. This would be consistent with the higher temperature required for the $NiFe_2O_4$ crystallite formation and growth observed in the analysis of the *in situ*

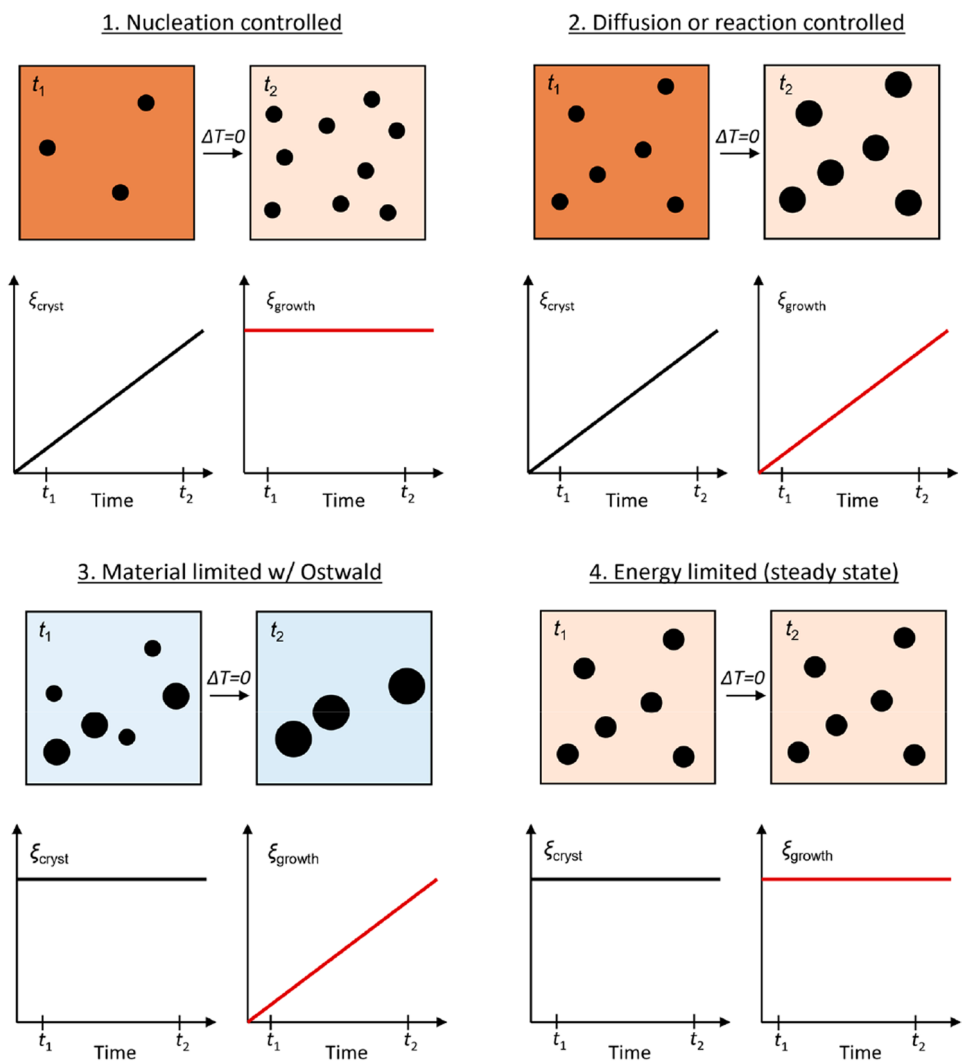

Figure 5. Illustration of the change in system composition between two reaction times ($t_1$ and $t_2$) under isothermal conditions along with the corresponding characteristic time-resolved evolution in extent of crystallization ($\xi_{cryst} = s/s_{final}$) and extent of growth ($\xi_{growth} = \langle V \rangle/\langle V \rangle_{final}$) for the specific limiting/controlling factors. Description in text.

PXRD data (discussed later) compared to the other spinel ferrites.

The long-range spinel structure is formed rapidly in under 40 s for all four compounds, and the nucleation seemingly happens over only a few data frames (10−15 s). Similar to the 150 °C synthesis of $CoFe_2O_4$, as the reaction progresses, the features sharpen, and the intensity of the $M_O$−$M_T$ peak (∼3.5 Å) increases relative to the neighboring $M_O$−$M_O$ peak (∼3 Å) indicating an equivalent nucleation and growth mechanism (see Figure 3b). After 120 s, all four compositions exhibit coherent structural order beyond 15 Å, however, the PDF for the $NiFe_2O_4$ sample show much lower peak damping at higher $r$, which is indicative of longer-range order. This agrees with the relative crystallite sizes observed in the *in situ* PXRD data (discussed later), where $NiFe_2O_4$ forms the largest crystallites (∼30 nm) at 250 °C.

**Size Control, Crystallization Dynamics, and Growth Kinetics—*In Situ* PXRD.** In addition to the nucleation and early stage formation/growth mechanisms discussed above, the extended hydrothermal crystallization and growth behavior of the $MFe_2O_4$ nanocrystallites were also investigated by *in situ* synchrotron PXRD. The *in situ* PXRD experiments were carried out on precursors prepared via the NaOH route as described in the experimental section. Data sets were collected at several different reaction temperatures for each spinel ferrite composition to study the effect of synthesis temperature on the crystallization rate and crystallite size evolution. Contour plots of the *in situ* synchrotron PXRD data can be found in the Supporting Information.

**Crystallization Dynamics—Qualitative Analytical Framework.** Using sequential Rietveld analysis of the collected *in situ* PXRD data, time-resolved values for the scale factor and mean isotropic crystallite volume, $\langle V \rangle = (4/3)\pi(\langle D \rangle/2)^3$, can be extracted. The scale factor can be used as an indicator of the extent of crystallization ($\xi_{cryst}$), as it is proportional to the total diffracting volume (and crystalline weight fraction) of the associated crystalline phase. Similarly, the extent of crystallite growth ($\xi_{growth}$) is proportional to $\langle V \rangle$. Specifically, the $\xi_{cryst}$ parameter, i.e., change in amount/volume of crystalline material, is equivalent to the refined scale factor normalized by its final stable value ($\xi_{cryst} = s/s_{final}$), while $\xi_{growth}$ is equivalent to the mean crystallite volume assuming isotropic morphology normalized by its final stable value ($\xi_{growth} = \langle V \rangle/\langle V \rangle_{final}$) under isothermal conditions. The combined evolution of these two parameters can be used to qualitatively evaluate the controlling/limiting factors for crystallization and growth as illustrated in Figure 5 and described in the following.

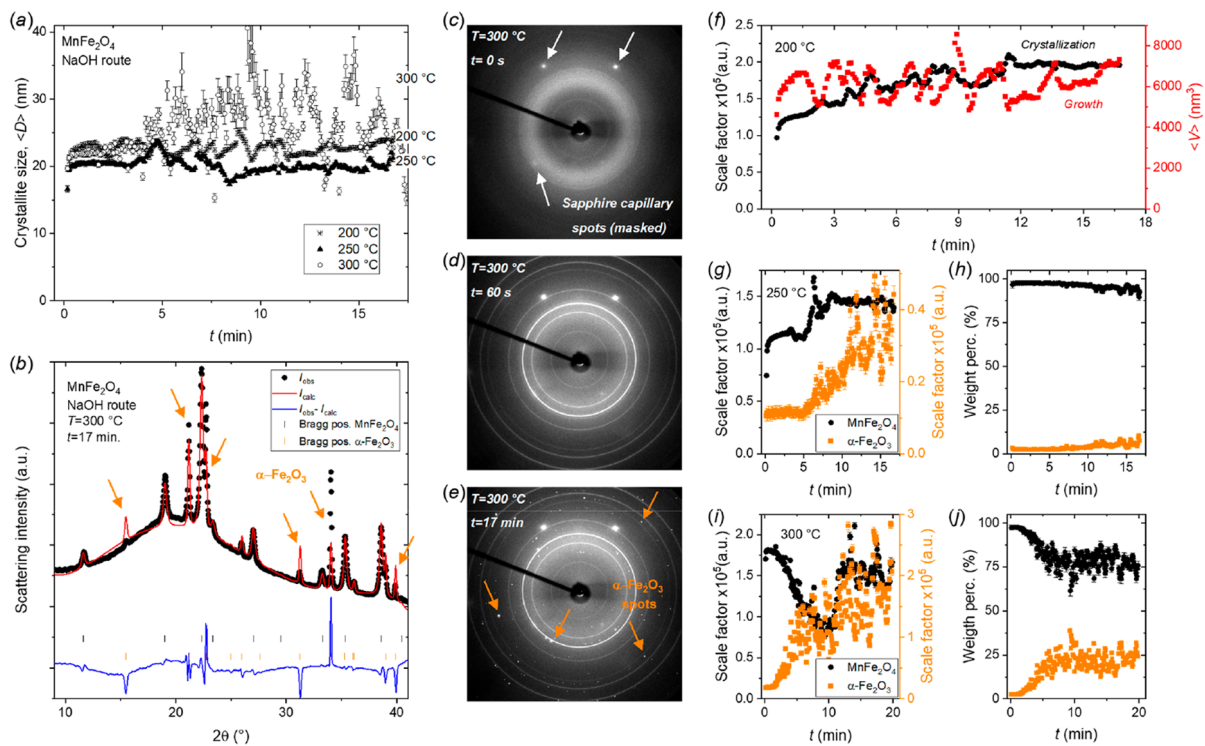

Figure 6. (a) Refined mean crystallite dimensions, ⟨$D$⟩, of MnFe$_2$O$_4$ as a function of hydrothermal reaction time (NaOH route) at the indicated temperatures. (b) PXRD pattern and Rietveld fit obtained after 17 min at 300 °C. The orange arrows indicate the poorly fitted peaks of the $\alpha$-Fe$_2$O$_3$ secondary phase due to its grainy nature. (c−e) 2D scattering data collected after the indicated reaction times at 300 °C. The orange arrows in (e) highlight a few of the many single-crystal spots from the $\alpha$-Fe$_2$O$_3$ secondary phase. (f) Refined scale factor (crystallization) and mean isotropic crystallite volume (growth) of MnFe$_2$O$_4$ as a function of reaction time for the 200 °C experiment. (g) Refined scale factors for the MnFe$_2$O$_4$ and the $\alpha$-Fe$_2$O$_3$ phases as a function of reaction time for the 250 °C experiment. (h) Refined weight fractions as a function of reaction time for the 250 °C experiment. (i) Refined scale factors of the MnFe$_2$O$_4$ and the $\alpha$-Fe$_2$O$_3$ phases as a function of reaction time for the 300 °C experiment. (j) Refined weight fractions as a function of reaction time for the 300 °C experiment.

For instantaneous application of a constant temperature (isothermal system, $\Delta T = 0$), we can assume the subsequent crystallization and/or crystallite growth to be controlled/limited by one or more of the following factors:

*1. Nucleation Controlled.* In this case, the kinetic energy barrier for nucleation is overcome while the barrier for incorporation of further precursor material into existing grains through surface reactions is either not overcome or is very limited above a specific equilibrium crystallite size. Consequently, crystallization is controlled/limited by the formation of new primary nuclei, while the mean crystallite size remains largely constant with time.

*2. Diffusion or Reaction Controlled.* The precursor concentration is far below the critical supersaturation level necessary to overcome the kinetic energy barrier for formation of new nuclei. Instead, the remaining precursor gradually crystallizes onto existing grains, leading to crystallization and growth by incorporation of material directly from the solution. In this case, the crystallization rate and, in turn, the growth is controlled either by the local supply of precursor material (diffusion limited) or by the energy barrier for incorporating atoms into the crystal structure at the exposed surface (reaction limited) through zero-order, first-order, or phase boundary reactions. Consequently, the mean crystallite volume will be observed to be directly proportional to (and gradually increase in tandem with) the amount/volume of crystalline material.

*3. Material Limited with Ostwald Ripening.* The energy barriers for both nucleation and growth are far exceeded and, as such, all precursor material is immediately precipitated in a burst of nucleation of many small crystallites. As no primary precursor material is left in solution, any subsequent growth will instead take place through Ostwald ripening, which involves a gradual dissolution of the smallest crystallites and recrystallization of the material onto larger ones. Therefore, the average crystallite size increases as a function of time, while the total amount of crystalline material remains constant.

*4. Energy Limited (Steady State).* The amount of energy in the system is insufficient to overcome the kinetic barrier for any of the crystallization and growth processes mentioned above (nucleation, surface reaction/incorporation, Ostwald ripening, etc.). Consequently, neither crystallization nor further growth takes place.

Notably, in most systems, the crystallization and growth during synthesis will be governed by a combination of several different factors at the various stages of the reaction.

In the following sections, the framework above will be used to evaluate the limiting factors for crystallization and growth in the studied systems. However, for several of the *in situ* experiments a final stable equilibrium was not attained within the time frame of experiment. Consequently, in order to avoid misrepresentation, the analysis is based on the non-normalized curves with absolute obtained values which are proportional to the extent of reaction (crystallization and growth) curves.

**MnFe$_2$O$_4$—Crystallization and Growth Mechanisms.** Figure 6a shows the evolution in the refined mean crystallite dimensions $\langle D \rangle$ as a function of heating time for the hydrothermal synthesis of MnFe$_2$O$_4$ nanoparticles by the NaOH route at 200, 250, and 300 °C. At all three synthesis temperatures, the MnFe$_2$O$_4$ crystallites grow to ∼20−25 nm in <30 s and remain in this range for the rest of the experiments (∼17.5 min.) with only minor variations in the refined sizes due to inherent fluctuations in experimental conditions. Consequently, the applied synthesis temperature has little-to-no impact on the obtained mean nanocrystallite size. Notably, the refined sizes and corresponding uncertainties of the 300 °C experiment vary drastically for heating times beyond approximately 5 min. This stems from a worsening of the Rietveld fit quality that is caused by the increasing scattering contribution from large α-Fe$_2$O$_3$ grains that are gradually formed in the reaction. The α-Fe$_2$O$_3$ structure have diffraction peaks overlapping with several of the broad MnFe$_2$O$_4$ nanocrystallite reflections, including the main MnFe$_2$O$_4$ (311) peak at a 2$\theta$ of ∼22.3° that overlaps with the (110) α-Fe$_2$O$_3$ reflection at ∼22.8° (see Figure 6b). As a result of the non-powder-like "macrocrystalline" nature of the α-Fe$_2$O$_3$ part of the sample, the absolute and relative intensities of the α-Fe$_2$O$_3$ Bragg reflections vary considerably between frames as the large grains rotate in an out of diffraction conditions in the beam. This prevents the scattering from α-Fe$_2$O$_3$ from being consistently and accurately fitted/described by incorporating this phase in the Rietveld analysis. This can be seen in Figure 6b, where the diffraction pattern obtained after 17 min at 300 °C has been fitted by a model containing both the spinel MnFe$_2$O$_4$ and the α-Fe$_2$O$_3$ phases. Figure 6c−e shows the 2D diffraction patterns collected on the area detector at three different times during the 300 °C experiment. At $t$ = 0 s, only the characteristic diffuse hydroxide ring from the precursor along with a few intense single-crystal sapphire spots from the capillary (which are masked out during data reduction) are observed (see Figure 6c). After $t$ = 60 s, the characteristic Debye−Scherrer ring powder pattern of the spinel structure has appeared and is the only crystalline phase visible in the scattering data besides the single-crystal spots from the sapphire capillary (see Figure 6d). However, at $t$ = 17 min (1020 s), a large number of distinct α-Fe$_2$O$_3$ single-crystal diffraction spots are clearly seen in the 2D data along with the spinel ferrite PXRD pattern. A few of these peaks have been marked with orange arrows in Figure 6e, although many more can be observed.

The formation of the iron oxide secondary phase seemingly depends on the applied synthesis temperature. At 200 °C, MnFe$_2$O$_4$ is the only crystalline phase observed in the *in situ* PXRD data over the entire experiment (see Supporting Information). Figure 6f (black symbols) shows the evolution of the refined scale factor for the MnFe$_2$O$_4$ phase in the 200 °C. As discussed earlier, the scale factors are proportional to the extent of crystallization ($\xi_{cryst}$), i.e., the relative change in the amount/volume of crystalline material, for the given phase. As such, in the 200 °C experiment, it is observed how approximately 50% of the MnFe$_2$O$_4$ formed during the experiment crystallizes within the initial few seconds of heating. Subsequently, additional material gradually crystallizes and the crystallization reaches an equilibrium level after approximately 11−12 min. Since the scale factor for a given phase is proportional to its crystalline volume fraction, this value can be compared to the mean isotropic crystallite volume ($\langle V \rangle$), which is proportional to the extent of growth ($\xi_{growth}$) discussed earlier to obtain information about the controlling/limiting crystallization and growth factors at play. While the refinement of crystallite size fluctuates drastically, as evident from the ∼30% variations in $\langle V \rangle$ (see Figure 6f, red symbols), the MnFe$_2$O$_4$ nanocrystallites seem to reach their equilibrium size after only a few seconds of starting the reaction at 200 °C. This indicates nucleation to initially be the controlling factor at this lower reaction temperature (reaction time <10 min), followed by steady state/equilibrium conditions (reaction time >10 min).

In the 250 °C experiment, the α-Fe$_2$O$_3$ secondary phase begins to crystallize as large grains after ∼5 min of heating (see Figure 6g). Only a minor amount of α-Fe$_2$O$_3$ is formed in the duration of the experiment constituting <10 wt % of the total crystalline material after 17.5 min of heating (see Figure 6h). Notably, the very small final amount of α-Fe$_2$O$_3$ is also responsible for the scale factor curve (Figure 6g) starting at a level of 25% of the final value. A minor amount of the phase is inherently (yet likely erroneously) considered present by the refinement model despite it not yet being formed (typically fitted into the background), and thus when the normalization denominator ($s_{final}$) is small the initial normalized scale factor value becomes artificially high despite no α-Fe$_2$O$_3$ being present. In the MnFe$_2$O$_4$ scale factor curve shown in Figure 6g, a discontinuous step is observed after ∼5 min of heating. The fact that a similar feature is also present in the α-Fe$_2$O$_3$ curve suggests this is likely related to sample movement.[70] Gas formation, capillary clogging/unclogging, temperature gradients, and other turbulence inducing factors (typically more common at higher reactions temperatures) may lead to sample movement, i.e., sudden changes in the amount of sample probed, and, in turn, peculiar variations in the refined scale factor. However, unless major movements take place bringing precursor/sample with different thermal history into the beam, the trends and values for refined weight fractions, crystallite sizes and atomic structure parameters remain largely unaffected by this, as it only results in a simple change in probed sample volume and thus signal scaling (signal-to-noise ratio).[61] Unfortunately, it does make it more difficult to analyze the progression of the scale factor curve. Discounting the jump caused by sample movement in the 250 °C experiment (at ∼5 min) and considering the quick equilibration of the crystallite size (see Figure 6a), it seems that all MnFe$_2$O$_4$ is immediately crystallized reaching a semisteady state, where MnFe$_2$O$_4$ nanocrystallites are subsequently being consumed to form α-Fe$_2$O$_3$, rather than the secondary α-Fe$_2$O$_3$ phase being formed out of solution.

While the 250 °C data by itself is inconclusive on this matter, the hypothesis above is further supported by the observations in the high temperature 300 °C experiment. Here, all MnFe$_2$O$_4$ is immediately crystallized (see Figure 6i), and the crystallites quickly grow to a 20−25 nm equilibrium size (see Figure 6a). A scale factor jump related to sample movement can be observed after ∼12 min of heating (see Figure 6i). However, observing the evolution in the scale factor curves for MnFe$_2$O$_4$ and α-Fe$_2$O$_3$ in the initial 10 min of the experiment, it is clear how MnFe$_2$O$_4$ is gradually consumed to form α-Fe$_2$O$_3$. Furthermore, as seen in Figure 6j, the phase composition of the system stabilizes after 5−8 min of reaction, with 20−25 wt % α-Fe$_2$O$_3$ and 75−80 wt % MnFe$_2$O$_4$.

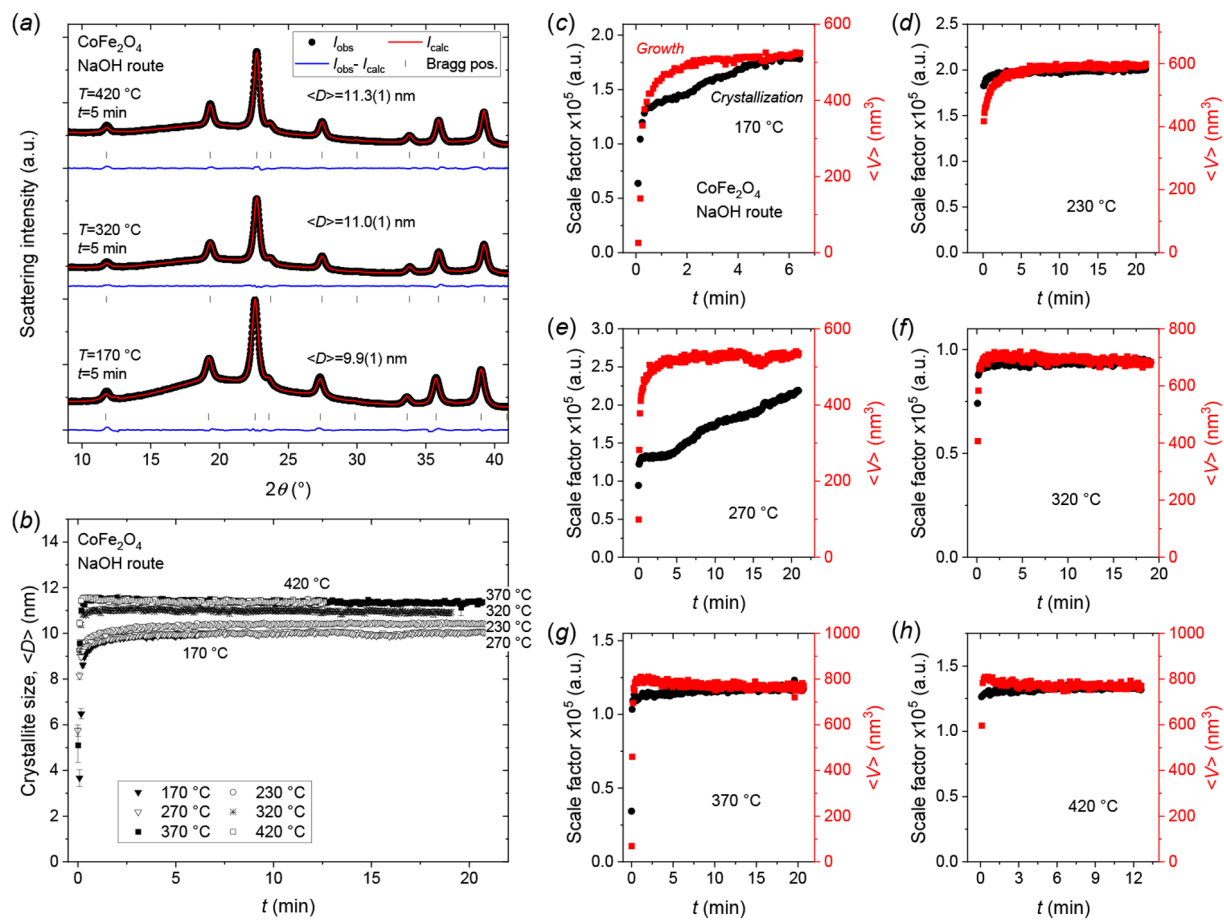

**Figure 7.** (a) Representative *in situ* PXRD patterns and Rietveld fits for CoFe$_2$O$_4$ nanocrystallites obtained after 5 min of hydrothermal treatment at the indicated temperatures. (b) Mean refined crystallite dimension as a function of hydrothermal reaction time for CoFe$_2$O$_4$ nanocrystallites at the indicated temperatures. (c–h) Refined scale factors (crystallization) and mean isotropic crystallite volumes (growth) as a function of reaction time for hydrothermal CoFe$_2$O$_4$ nanoparticle syntheses conducted at (c) 170 °C, (d) 230 °C, (e) 270 °C, (f) 320 °C, (g) 370 °C, and (h) 420 °C.

Notably, the conversion of the primary MnFe$_2$O$_4$ phase to α-Fe$_2$O$_3$ occurs earlier and more rapidly at higher synthesis temperatures.

The observed formation of α-Fe$_2$O$_3$ begs the question of what happens to the corresponding Mn-content as Fe$^{3+}$ is removed from the spinel phase. Do the Mn-ions go back into solution? Is an amorphous manganese oxide phase being concurrently formed? Or does the spinel phase become increasingly Mn-rich with charge balance being preserved through Mn$^{2+}$ oxidation to Mn$^{3+}$? To answer this, further studies are necessary as the present *in situ* PXRD data does not provide any conclusive evidence.

**CoFe$_2$O$_4$—Crystallization and Growth Mechanisms.** For all the CoFe$_2$O$_4$ experiments, the intended nanosized CoFe$_2$O$_4$ crystallites were the only crystalline product observed in the *in situ* PXRD data. Figure 7a shows representative *in situ* PXRD patterns obtained after 5 min in the 170, 320, and 420 °C experiments. As evident from the very similar diffraction patterns and peak profiles, only very limited variations in crystallite size were observed when varying the synthesis temperature. The evolution in the refined mean CoFe$_2$O$_4$ crystallite dimensions ⟨D⟩ as a function of heating time at subcritical (170, 230, 270, and 320 °C), near-critical (370 °C) and supercritical (420 °C) hydrothermal conditions is shown in Figure 7b. Despite the 250 °C difference between the lowest (170 °C) and highest (420 °C) applied temperatures, little-to-no change is observed in the resulting mean crystallite sizes, which all quickly equilibrate in the ∼10−12 nm range. This is in agreement with our previous *in situ* PXRD study of CoFe$_2$O$_4$ nanoparticle synthesis, where we found that changing the concentration of the metal salt solution at the moment of NaOH addition during the precursor preparation, provides a better handle (compared to reaction temperature) for tuning the resulting CoFe$_2$O$_4$ crystallite sizes in this system.[71]

Figure 7c–h show the evolution of the scale factors and mean isotropic crystallite volume, ⟨V⟩, at the different reaction temperatures. At 170 °C, the growth reaches an equilibrium value relatively quickly (<2.5 min) while crystallization gradually increases throughout the entire experiment (7 min). This behavior would indicate a mixed nucleation- and reaction/diffusion-controlled growth at the early stages, which transitions to a mostly nucleation-controlled mechanism later. At the early stages of the 230 °C experiment, i.e., initial ∼120 s, a slight divergence of the two curves is observed with the crystallization saturating faster than the growth (see Figure 7d). This indicates that the reaction is material-limited, and that some growth is occurring by Ostwald ripening in this region. Interestingly, the 270 °C experiment exhibits opposite

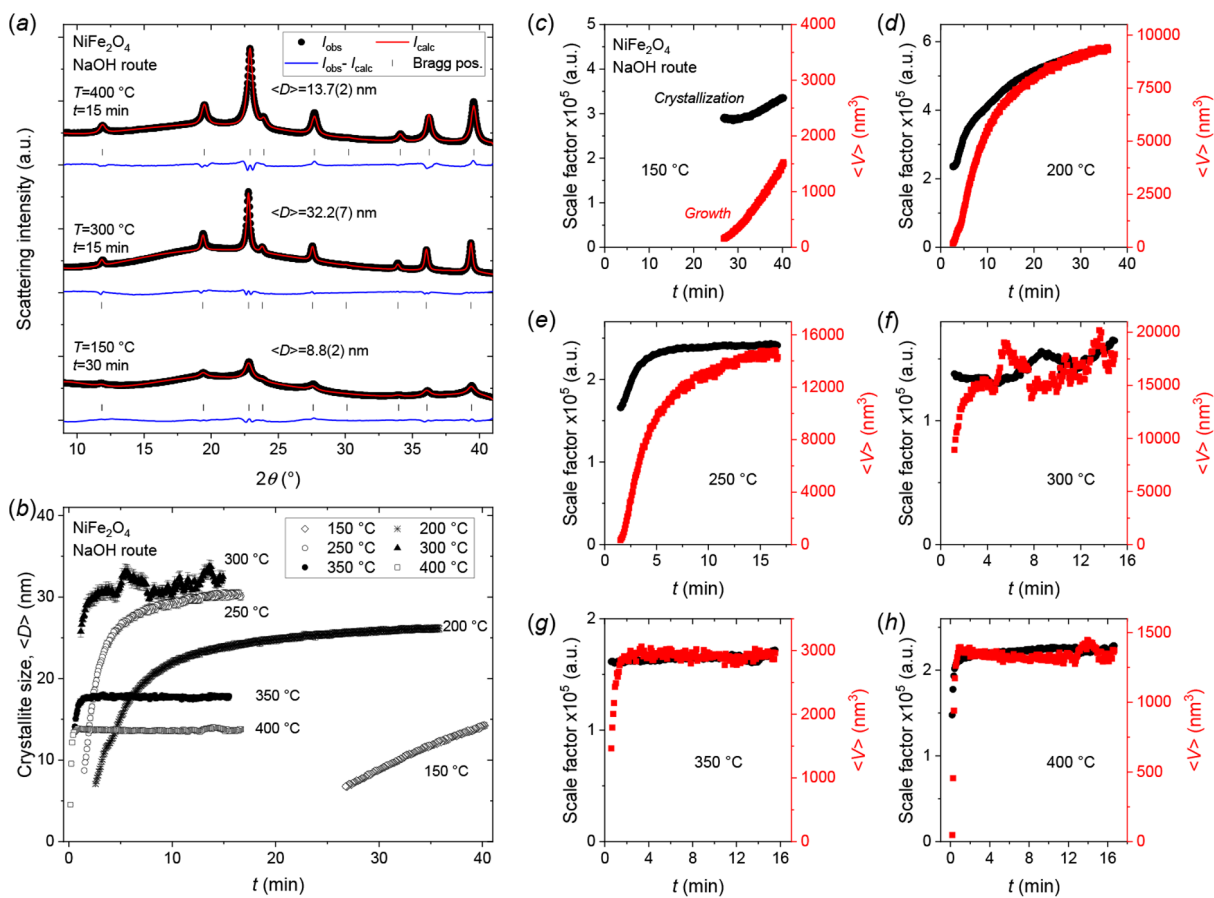

**Figure 8.** (a) Representative *in situ* PXRD patterns and Rietveld fits for NiFe$_2$O$_4$ nanocrystallites obtained after 30 min of hydrothermal treatment at 150 °C (bottom), 15 min at 300 °C (middle) and 15 min at 400 °C (top), respectively. (b) Evolution in refined mean crystallite dimensions, ⟨D⟩, for the NiFe$_2$O$_4$ syntheses at the indicated temperatures. (c−h) Double-Y plots of refined scale factors (crystallization) and mean isotropic crystallite volumes (growth) as a function of reaction time for hydrothermal NiFe$_2$O$_4$ nanoparticle syntheses conducted at (c) 150 °C, (d) 200 °C, (e) 250 °C, (f) 300 °C, (g) 350 °C, and (h) 400 °C.

behavior (see Figure 7e) with a stable mean nanocrystallite volume being relatively rapidly achieved (<3 min) while the crystallization gradually increases without reaching steady state within the time of the experiment (∼21 min). Since ⟨V⟩ does not increase, this must mean that the additional diffraction signal comes from new crystalline material that is formed by continuous nucleation of new nuclei that rapidly grow to the equilibrium size (∼10 nm). The 270 °C trends are similar to those observed for 170 °C, but at an approximately 4 times faster rate. At higher temperatures (320, 370, and 420 °C), full crystallization is achieved rapidly in <20 s and stable equilibrium crystallite sizes are almost instantly obtained. Consequently, at temperatures above 320 °C the critical energy barrier for nucleation must be far exceeded. This steady state with no further growth by Ostwald ripening, indicates that the formed CoFe$_2$O$_4$ crystallites are both kinetically and thermodynamically stable at the given conditions despite their relatively moderate sizes of 10−12 nm.

**NiFe$_2$O$_4$—Crystallization and Growth Mechanisms.** *In situ* synchrotron PXRD experiments were carried out for the hydrothermal synthesis of NiFe$_2$O$_4$ nanocrystallites at six different temperatures: four in the subcritical (150, 200, 250, 300 °C), one in the near-critical (350 °C) and one in the supercritical (400 °C) regime. Notably, the *in situ* PXRD data of all the experiments in the series showed NiFe$_2$O$_4$ as the only crystalline product (see Figure 8a). Interestingly, despite equivalent synthesis procedure and parameters being employed (i.e., the only difference being the divalent transition metal ion type/source, i.e., 2.0 M Ni(NO$_3$)$_2$·6H$_2$O vs Co(NO$_3$)$_2$·6H$_2$O), the crystallization and growth behavior observed in the NiFe$_2$O$_4$ system is considerably different compared to that of the CoFe$_2$O$_4$. Here, varying the reaction temperature is found to considerably change the resulting nanocrystallite sizes as shown in Figure 8a,b. Furthermore, a more complex relationship between applied reaction temperature and the growth behavior/mechanism is observed. In Figure 8a, PXRD patterns obtained at three different reaction temperatures, i.e., 150, 300, and 400 °C, after 30, 15, and 15 min of hydrothermal treatment, respectively, are shown, illustrating the large differences in the obtained nano-/microstructure of the NiFe$_2$O$_4$ products. At the lowest temperature (150 °C), after 30 min of hydrothermal treatment a mean crystallite size of 8.8(2) nm is obtained. Interestingly, the largest crystallite size (32.2(7) nm after 15 min) is observed at the intermediate reaction temperature (300 °C), while a clearly lower crystallite size (13.7(2) nm after 15 min) is obtained at the highest applied temperature (400 °C). Inspecting the evolution in crystallite size as a function of reaction time, shown in Figure 8b, along with the scale factor (crystallization) and mean isotropic crystallite volume (growth) curves in Figure 8c−h, reveals how increasing the synthesis temperature causes both the crystallization and growth rate to increase, however, at temperatures above 300 °C the resulting equilibrium mean crystallite size decreases (see Figure 8b).

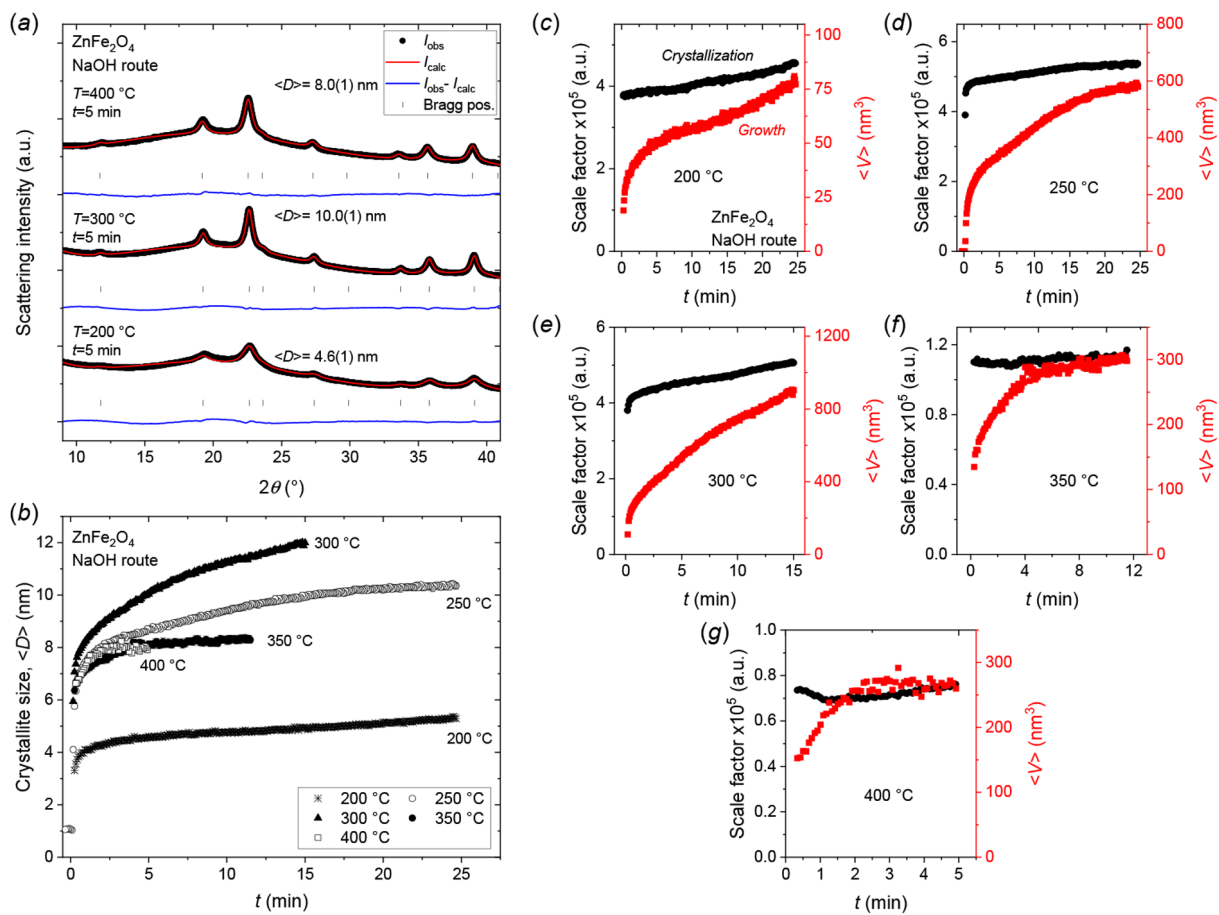

**Figure 9.** (a) Representative *in situ* PXRD patterns and Rietveld fits for ZnFe$_2$O$_4$ nanocrystallites obtained after 5 min of hydrothermal treatment of the NaOH route precursor at the indicated temperatures. (b) Mean refined ZnFe$_2$O$_4$ crystallite dimension as a function of reaction time at the indicated temperatures. (c–g) Refined scale factors (crystallization) and mean isotropic crystallite volumes (growth) as a function of reaction time for hydrothermal ZnFe$_2$O$_4$ nanoparticle syntheses conducted at (c) 200 °C, (d) 250 °C, (e) 300 °C, (f) 350 °C, and (g) 400 °C.

This growth behavior and variation in equilibrium size are likely the result of temperature-dependent contributions from several controlling/limiting mechanisms. In the subcritical range (∼100−300 °C, 250 bar), the solvent properties of water remain largely constant, meaning that the extra energy added by increasing the synthesis temperature simply induces faster crystallization and growth. However, at near- (350 °C, 250 bar) and supercritical (400 °C, 250 bar) hydrothermal conditions the solvent properties of water, including factors such as the dielectric constant, density and ion dissociation constant, change considerably, and it becomes similar to nonpolar hexane thereby acting as an antisolvent forcing the precipitation.[72,73] Consequently, when heating rapidly to near- or supercritical conditions, a burst of nucleation is induced, where a much larger amount of primary particles are formed, which results in smaller average crystallite sizes and narrower size distributions.

At 150 °C (see Figure 8c), the crystallization and growth slowly commence after an incubation time of ∼25 min and clearly does not finish within the time of the experiment (∼40 min). As such, the scale factor and ⟨V⟩ curves shown in Figure 8c do not reach a final steady state value and should be evaluated with care. However, the increase in both crystallization and growth curves indicates that the reaction is at least in part diffusion or reaction-controlled.

At 200 °C (see Figure 8d), the crystallization and growth curves initially (0−10 min) increase in parallel and later (>10 min) gradually converge and equilibrate. This indicates that the reaction is controlled by a combination of nucleation and growth by diffusion at the early stages, with the nucleation contribution diminishing with extended reaction time as the diffusion/reaction-controlled growth continues. The crystallization and growth curves do not fully reach steady state within the time of the experiment (∼35 min), but a mean crystallite size of 26.1(4) nm is obtained from the refinement of the last pattern in the experiment.

For the 250 °C experiment (see Figure 8e), the crystallization reaches steady state much faster (<7.5 min) compared to the crystallite growth, which does not fully equilibrate within the time of the experiment (∼17 min). The crystallization and growth mechanisms again gradually transition from being initially governed mainly by nucleation (<2.5 min), toward being more diffusion/reaction controlled at the intermediate stage (∼2.5−7.5 min), and finally exhibiting Ostwald ripening at extended reaction times (>7.5 min). A mean crystallite size of 30.1(4) nm is obtained at the end of the experiment following ~17 min of heating at 250 °C.

The evolution in crystallization and growth for the 300 °C experiment is relatively unstable (see Figure 8f), likely due to sample movement and/or fluctuations in beam intensity. Consequently, care must be taken when interpreting the data. However, steady state is seemingly achieved for both crystallization and growth within the initial 5 min. Notably, this experiment yields the largest crystallites within the studied temperature/time parameters attaining a mean size of 32.2(7) nm after 15 min of hydrothermal treatment at 300 °C.

Similar observations can be made for the 350 and 400 °C NiFe$_2$O$_4$ experiments, where steady state for both crystallization and growth is attained even faster, i.e., in <90 s and in <45 s, respectively. Interestingly, as discussed earlier, the equilibrium crystallite sizes are found to decrease compared to the lower reaction temperatures with mean sizes of 17.8(3) nm and 13.7(2) nm being obtained after 16 min in the 350 and 400 °C experiments, respectively. Furthermore, Ostwald ripening is observed in the 250 and 300 °C for relatively large crystallites (>20 nm) while no Ostwald ripening is seen in the higher temperature 350 and 400 °C despite the crystallites being smaller. This trend may initially seem counterintuitive but may be explained by the various mechanisms at play and taking into consideration the effect of crystallite size distributions (see discussion in Supporting Information).

**ZnFe$_2$O$_4$—Crystallization and Growth Mechanisms.** The hydrothermal formation of ZnFe$_2$O$_4$ nanocrystallites (NaOH route) was investigated by *in situ* synchrotron PXRD at five different temperatures, three at subcritical conditions (200, 250, 300 °C), one at near-critical condition (350 °C) and one in the supercritical (400 °C) regime. For all experiments, only diffraction peaks associated with the characteristic spinel ferrite structure were observed (see Figure 9a) indicating that ZnFe$_2$O$_4$ is the only crystalline phase formed. Again, the crystallization and growth behaviors are considerably different from those of the Mn-, Co-, and Ni-ferrites discussed earlier. Similar to the NiFe$_2$O$_4$ experiments, changing the applied temperature leads to variations in the resulting crystallite size, albeit attaining much smaller sizes (4–12 nm) at equivalent temperatures (see Figure 9b). As for NiFe$_2$O$_4$, the largest crystallite sizes (~10–12 nm) are observed for the samples prepared at intermediate temperature (300 °C), with the resulting size being smaller (~8 nm) when the temperature is increased to the near- and supercritical range (350, 400 °C). Albeit, the smallest crystallites (~4–5 nm) are, however, still obtained at the lowest applied temperature of 200 °C. As discussed earlier for NiFe$_2$O$_4$, this behavior is likely related to the changes in the solvent's properties when heating close to and above its critical point (374 °C, 221 bar), where precipitation suddenly becomes vastly favored.[73] This leads to formation of a much larger number of primary particles and, in turn, a smaller average crystallite size and narrower size distribution.

Figure 9c–g show the relative evolution in crystallization and crystallite growth at the examined synthesis temperatures. At the three lowest temperatures (Figure 9c–e), neither the crystallization nor the growth reaches steady state within the timespans of the experiments. The curves could, considering the seemingly slower increase in crystallization compared to growth, indicate a mostly diffusion/reaction-controlled growth with a contribution from Ostwald ripening. At 350 °C, full crystallization is achieved almost immediately, while growth continues by Ostwald ripening for 4–5 min before steady state is reached (see Figure 9f). At 400 °C, the crystallization immediately reaches steady state while the growth curve takes ~2 min to equilibrate (see Figure 9g). Notably, the absence of Ostwald ripening despite the relatively small crystallite size and high applied temperature may be attributed to combination of size distribution effects and the inherent volume-weighting of the sizes from the PXRD experiment (see discussion in Supporting Information).

## DISCUSSION

Solvothermal nanoparticle formation has traditionally been discussed in terms of simplistic thermodynamic arguments without noteworthy consideration of the distinct chemical natures of the precursor species. However, in recent years studies have demonstrated how the local precursor structure plays an extremely important role in directing the formation of the material and how the classical LaMer-type nucleation theory often falls short in describing the complex early stage chemical mechanisms at play.[46] In the present study, *in situ* TS with PDF analysis (NH$_4$OH route) has been used to elucidate the hydrothermal nucleation mechanism for the four spinel ferrite nanoparticle systems, MnFe$_2$O$_4$, CoFe$_2$O$_4$, NiFe$_2$O$_4$ and ZnFe$_2$O$_4$. From the PDF analysis, we demonstrate how the nucleation occurs via equivalent mechanisms for the four spinel ferrite compounds. As illustrated in Figure 3c, the TMs initially form edge-sharing octahedrally coordinated hydroxide units (monomers/dimers and in some cases trimers) in the aqueous precursor, which upon hydrothermal treatment nucleate through linking by tetrahedrally coordinated TMs. As discussed earlier, the broad features from the precrystalline clusters observed in the early stage PDFs can often be equally well described by the local structure of a variety of metal oxide clusters. As such, determining a meaningful reaction pathway is often best done working backward starting from the PDFs of the crystalline product, with the interpretation being informed by the crystalline structure determined from PXRD analysis (or known structures from literature), and letting the relative changes in PDF peak positions and intensities help elucidate the mechanism. The complementarity and synergy of the two techniques are illustrated in Figure 1c and the deductive reasoning behind the concluded mechanism is explained earlier in the PDF section of the Results.

Interestingly, the *in situ* PXRD data (NaOH route) reveals very different crystallization and growth behaviors of the studied compounds. For NiFe$_2$O$_4$ and ZnFe$_2$O$_4$, the crystallites grow gradually at lower temperatures (<250 °C) and show a reduction in final mean crystallite size at higher temperatures (>350 °C) in accordance with LaMer theory (burst of nucleation). Meanwhile, for all investigated reaction temperatures, the MnFe$_2$O$_4$ (200–300 °C) and CoFe$_2$O$_4$ (170–420 °C) nanocrystallites almost instantaneously grow to equilibrium sizes of 20–25 nm and 10–12 nm, respectively. These drastic differences in the growth behavior (NaOH route), which are solely caused by changing the divalent ion (Mn/Co/Ni/Zn), points to an underlying chemical mechanism. Notably, our previous work on the related hydrothermally prepared spinel-structured iron oxides (magnetite Fe$_3$O$_4$, maghemite γ-Fe$_2$O$_3$) revealed a similar mechanism for nucleation to the ones observed for the spinel ferrites studied here.[69,74,75] However, while relatively simple structural models were used for the *in situ* data analysis, detailed *ex situ* characterization has revealed complex crystal- and local structures with vacancy formation and ordering (symmetry lowering),[76] as well as core–shell formation[77,78] or stoichiometric gradients in the spinel ferrite nanoparticles.[76] Similarly, in another spinel-structured nanoparticle system, $ZnAl_2O_4$, recent work using *in-* and *ex situ* characterization has highlighted the importance of not only considering the metal atom inversion between octahedral and tetrahedral sites, but also other defects such as interstitial atoms, which can significantly influence the physical properties.[79,80] Consequently, complementary use of *in-* an *ex situ* methods will often be necessary to fully elucidate the involved mechanisms. In this context, our previous *ex situ* structural studies of hydrothermally prepared spinel ferrite nanocrystallites using neutron powder diffraction (NPD), which opposed to X-ray scattering provides scattering contrast between the neighboring transition metals, have shown the products to be stoichiometric $MFe_2O_4$, but revealed site preferences different to the conventional bulk equivalents.[68,70,81,82] The nanosized $MnFe_2O_4$ and $CoFe_2O_4$ crystallites were found to exhibit mostly random disordered spinel structures, while $NiFe_2O_4$ is a completely inverse spinel and $ZnFe_2O_4$ is semidisordered, close to a normal spinel. Here, it is interesting that the compounds that typically exhibit mixed spinel structures ($MnFe_2O_4$ and $CoFe_2O_4$) were found to grow more rapidly to specific equilibrium sizes, while the crystallite size of normal spinel $ZnFe_2O_4$ and inverse spinel $NiFe_2O_4$ evolve gradually at equivalent synthesis temperatures. The lack of preference of $Mn^{2+}$ or $Co^{2+}$ (relative to $Fe^{3+}$) for any of the two spinel sites may lower the barrier for nanocrystal nucleation and growth compared to $Zn^{2+}$ and $Ni^{2+}$, which have strong preferences for the tetrahedral 8a and octahedral 16d sites, respectively. This could indicate that the site preference determined by the chemical nature of the specific elements plays a key role for the nucleation and growth of spinel ferrite nanoparticles and potentially for other nanoparticle systems.

The evolution in refined lattice parameters (see Supporting Information) over the course of the reactions may potentially provide indications of whether any compositional changes or structural reconfigurations are occurring during crystallization and growth. However, for *in situ* PXRD experiments of the type carried out here, several sample characteristics (composition, cation inversion, crystallite size) and reaction parameters (temperature, heating rate, pressure) may affect the obtained unit cell parameters making it difficult to reliably determine the origin of any lattice changes. In particular, the lattice parameters tend to be highly influenced both by the effects of thermal expansion and crystallite size.[61] The rapid and efficient heating of the employed setup reduces the errors due to thermal lag to the first few data frames, however, the shifts in lattice parameters due to differences in absolute applied synthesis temperature between experiments must be considered. Furthermore, in ultrafine crystallites, a considerable fraction of the atoms will be situated at the surface where defects and variations due to interfacial effects occur resulting in differences in lattice parameters. However, as the crystallites grow, the surface-to-bulk ratio is reduced and the average cell length will tend toward the characteristic bulk value for the compound (typically lower due to tighter binding in the bulk). Therefore, while higher temperatures lead to thermal expansion and thus typically increased lattice parameters, this can be more than offset by a reduction in cell parameters due to crystallite growth. These opposing effects, in combination with potential changes in composition, complicate the analysis of the evolution in cell parameters. Consequently, for the studied syntheses no reliable conclusions about structural/compositional evolution during the crystallization can be made based on the refined lattice parameters.

As discussed in the beginning of the Results section, two different precursor preparation routes ($NH_4OH$ and NaOH routes) were used for the *in situ* PXRD and TS experiments, due to the different capillary type required for the TS experiments. This means that different precursor pH, counterions ($NO_3^-$ and $Cl^-$) and metal ion concentrations were used, which may affect the course of the hydrothermal reaction pathways. Thus, the observations from the TS and PXRD experiments may not be directly comparable. Indeed, in the case of $MnFe_2O_4$, using the highly concentrated NaOH favors precipitation of mixed valence $Mn_3O_4$ or trivalent $Mn(OH)_3$ rather than the desired divalent $Mn(OH)_2$ in highly alkaline aqueous conditions.[83,84] This is likely the reason for the observation of $\alpha$-$Fe_2O_3$ impurities when using the NaOH route due to the excess Fe present after nonstoichiometric $Mn_{1+x}Fe_{2-x}O_4$ formation. On the other hand, the pure $MnFe_2O_4$ spinel phase was observed to form when hydrothermally treating the precursor prepared using the $NH_4OH$ route. In this respect, Mn is likely the most problematic of the studied transition metals, as Fe precipitates as the trivalent oxyhydroxide (FeOOH) and Co, Ni and Zn tend to form divalent hydroxides ($M(OH)_2$) in alkaline aqueous conditions.[84,85] Furthermore, we have previously shown how hydrothermally treating the FeOOH precipitate alone (without the presence of divalent $M(OH)_2$ in the precursor) leads to formation of the thermodynamically preferred $\alpha$-$Fe_2O_3$ phase rather than the related spinel iron oxide phases $\gamma$-$Fe_2O_3$ or $Fe_3O_4$.[71] As such, while the kinetics of the reactions are likely affected by differences in metal ion concentration in the two precursor types (i.e., the resulting degree of supersaturation achieved at different temperatures), we expect the reaction mechanisms and end product phase of the $CoFe_2O_4$, $NiFe_2O_4$, and $ZnFe_2O_4$ systems to largely remain the same. This is further supported by our previous studies where spinel phase $CoFe_2O_4$ is consistently observed to crystallize when using different coprecipitation conditions, hydrothermal synthesis procedures and reaction parameters.[71,86]

## CONCLUSIONS

The nucleation, crystallization, and growth of spinel ferrite nanocrystallites have been studied by *in situ* synchrotron total scattering with PDF analysis ($NH_4OH$ route) and PXRD with sequential Rietveld modeling (NaOH route). The *in situ* TS experiments were carried out on 0.6 M TM hydroxide precursors prepared from aqueous metal chloride solutions using 24.5% $NH_4OH$ as the precipitating base. The hydrothermal nucleation of the spinel ferrite compounds under the studied conditions was found to take place from edge-sharing octahedral hydroxide units (monomers/dimers and in some cases trimers) in the precursor, which upon heating nucleate through linking via tetrahedrally coordinated TMs. The *in situ* PXRD experiments were carried out on 1.2 M TM hydroxide precursors prepared from aqueous metal nitrate solutions using 16 M NaOH as the precipitating base. Rietveld and peak profile analysis shows that at all investigated synthesis temperatures the $MnFe_2O_4$ (200–300 °C) and $CoFe_2O_4$ (230–420 °C) nanocrystallites rapidly grow to an equilibrium size of 20–25 nm and 10–12 nm, respectively, thus indicating limited possibility of targeted size control by variation of temperature. However, for NiFe$_2$O$_4$ (150–400 °C) and ZnFe$_2$O$_4$ (200–400 °C) the growth occurs gradually in the low temperature range allowing specific sizes to be targeted. In the intermediate range, the moderate nucleation and subsequent growth by diffusion allows the largest crystallites to be obtained, while at the highest temperature the burst of nucleation spends all precursor material, and subsequently only limited growth takes place by Ostwald ripening. Interestingly, it is the compounds typically exhibiting mixed spinel structures (MnFe$_2$O$_4$ and CoFe$_2$O$_4$) that grow rapidly to specific equilibrium sizes, while the crystallite size of normal spinel ZnFe$_2$O$_4$ and inverse spinel NiFe$_2$O$_4$ evolve gradually at equivalent synthesis temperatures. This appears to indicate that the lack of preference of Mn$^{2+}$ or Co$^{2+}$ (relative to Fe$^{3+}$) for any of the two spinel sites eases the nanocrystal nucleation and growth compared to the case of Zn$^{2+}$ and Ni$^{2+}$, which have strong preferences for the tetrahedral 8$a$ and octahedral 16$d$ sites, respectively. For CoFe$_2$O$_4$ and MnFe$_2$O$_4$ the kinetic crystallization barrier is greatly exceeded at all tested temperatures causing rapid full precipitation of spinel ferrite nanocrystallites. However, the MnFe$_2$O$_4$ particles are unstable at higher temperatures where they are gradually consumed to form the thermodynamically stable $\alpha$-Fe$_2$O$_3$ (hematite) phase. For ZnFe$_2$O$_4$ and NiFe$_2$O$_4$, it is observed how the growth of crystallites is often governed by a complex combination of limiting factors that vary throughout the crystallite growth thereby complicating the deconvolution of the contributions. While several questions remain, this extensive study provides detailed insight into the mechanisms at play during the hydrothermal formation and growth of spinel ferrite nanoparticles.

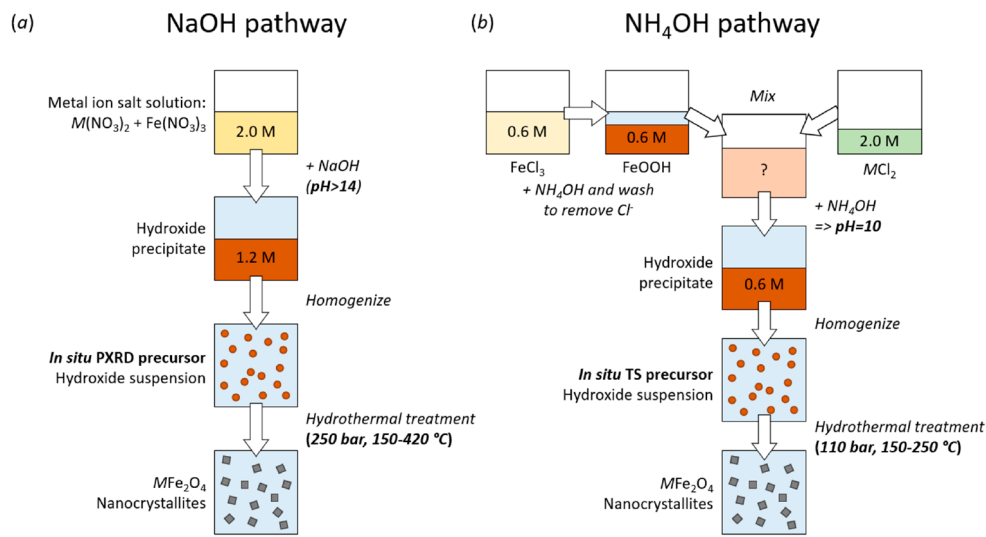

**Figure 10.** Schematic illustrations of the employed synthesis pathways. (a) The NaOH route used for the *in situ* PXRD experiments and (b) the NH$_4$OH route used for the *in situ* TS experiments.

## METHODS

**Precursor Preparation and Synthesis.** The precursors for the spinel ferrites were prepared by coprecipitation of trivalent iron and divalent transition metal hydroxides from aqueous salt solutions with the desired nominal stoichiometry. The coprecipitated precursors were then treated hydrothermally at elevated temperatures ($T$ = 150–420 °C) and pressures (110–250 bar) to induce formation of nanocrystallites. The two different precursor preparation routes, from here on referred to as the NaOH route and the NH$_4$OH route, are illustrated in Figure 10.

*Precursor Preparation—NaOH Route.* Precursor solutions of 2.0 M Fe(NO$_3$)$_3$·9H$_2$O and 2.0 M $M$(NO$_3$)$_2$·6H$_2$O ($M$ = Co, Ni, Zn), FeCl$_2$ or MnCl$_2$ (all chemicals being Sigma-Aldrich, ≥98% purity) were mixed in the stoichiometric ratio of the target compound. Subsequently, 16 M NaOH solution (equivalent to 1.25 times the molar amount of NO$_3^-$ ions) was added dropwise under constant magnetic stirring, leading to precipitation of a viscous hydroxide precipitate with pH > 14 and a final metal ion concentration of 1.2 M ($M^{2+}$ and Fe$^{3+}$). Notably, oxidation of Mn$^{2+}$ to Mn$^{3+}$ at high pH complicates the preparation of phase pure MnFe$_2$O$_4$ nanocrystallites via the NaOH route.

*Precursor Preparation—NH$_4$OH Route.* An aqueous 0.6 M iron(III) oxyhydroxide dispersion was prepared by dropwise addition of 24.5% NH$_4$OH to a solution of FeCl$_3$·6H$_2$O (Sigma-Aldrich, ≥98% purity) under constant magnetic stirring until a pH of 10 was reached. The FeOOH dispersion was repeatedly washed with demineralized water, centrifuged (3 min, 2000 rpm) and decanted until the supernatant pH was under 8. Subsequently, the prepared 0.6 M FeOOH dispersion was mixed with 2.0 M aqueous $M$Cl$_2$ solution ($M$ = Mn, Co, Ni, or Zn depending on the desired product) in the desired nominal molar amount. Then, 24.5% NH$_4$OH was added dropwise under constant magnetic stirring until a pH of 10 was reached giving a final metal ion concentration of 0.6 M in the precursor.

*In Situ Experimental Setup and Synthesis.* The *in situ* PXRD and TS experiments were carried out using the *in situ* setup illustrated in Figure 1b. Detailed descriptions of the setup and experimental procedure have previously been published.[61−63] The prepared precursors were loaded into a single-crystal sapphire capillary (inner diameter of 0.60 mm) for the *in situ* PXRD or a fused silica capillary (inner diameter of 0.70 mm) for the *in situ* total scattering experiments. Note that the more mechanically and chemically stable single-crystal sapphire capillaries employed for the *in situ* PXRD experiments cannot be used for the *in situ* TS experiments due to the excessive number of single-crystal spots, which would appear in the much larger $Q$-range of the TS data and severely complicate the data analysis. Instead, the amorphous fused silica capillaries, which produce an easily subtractable smooth amorphous background signal, are used. The capillaries were mounted in the setup and the system was pressurized with deionized water prior to initiating the heating using a HPLC pump connected via Swagelok fittings meaning that the initial scattering data at $t$ = 0 s were collected on the pressurized precursor. For the PXRD experiments a pressure of 250 bar was employed in order to access near-critical and supercritical hydrothermal conditions ($T > 374$ °C and $p > 221$ bar).[87] For the *in situ* TS a lower pressure of 110 bar was employed due to the tendency of the more flexible fused silica capillaries to bend and crack under high applied pressures. The *in situ* TS and PXRD experiments were carried out at reaction temperatures in the range 150–400 °C. The small sample volume along with the high air flow and efficiency of the heater ensure a very rapid heating of the system resulting in the temperature generally reaching >95% of the target within 10 s.[61] As mentioned, the NaOH route is generally preferred for the present *in situ* PXRD and TS experiments, as the higher precursor concentration increases the amount of scattering material probed. However, the highly concentrated and strongly alkaline 16 M NaOH solution corrodes the fused silica capillaries used for the *in situ* TS experiments, thereby making the use of the weaker base $NH_4OH$ necessary in this case. A detailed discussion about the various considerations, trade-offs, and pitfalls, when conducting *in situ* synchrotron PXRD studies of solvothermal nanoparticle synthesis has previously been published.[61]

**Characterization.** *In Situ X-ray Total Scattering.* The *in situ* total scattering experiments were carried out at the Powder Diffraction and Total Scattering beamline, P02.1, PETRA III, DESY, Hamburg, Germany.[88] The total scattering data was collected with a PerkinElmer XRD1621 amorphous silicon detector (2048 × 2048 $px^2$, pixel size 200 × 200 $\mu m^2$) with a sample-to-detector distance of 240 mm and a wavelength of 0.2072 Å (60 keV, $Q_{max} \approx 20$ Å$^{-1}$). The time resolution for the data collection was 5 s. The exact sample-to-detector distance and wavelength in the given experiments were calibrated using data collected from a NIST $LaB_6$ 660b PXRD standard reference material in the same instrumental configuration.

*In Situ Synchrotron Powder X-ray Diffraction.* The *in situ* synchrotron PXRD experiments were carried out over several experimental beamtimes at the Crystallography Beamline, I711, MAX-II, Lund, Sweden. The diffraction data was collected with an Oxford Diffraction Titan CCD area detector (diameter = 165 mm, pixel size 60 × 60 $\mu m^2$, 2 × 2 binning) with a sample-to-detector distance of ≈80–90 mm and a wavelength of ≈1.00 Å (12.4 keV, $Q_{max} \approx 4.3–5.0$). A time resolution of 5 s per 2D data set was attained using a 4 s exposure time and a detector readout time of 1 s. The exact sample-to-detector distance and wavelength in the given experiments for each individual beamtime were calibrated from data collected on a capillary loaded with NIST $LaB_6$ 660b PXRD standard reference material in the same instrumental configuration.

**Structural Analysis.** *In Situ TS Data Reduction and PDF Analysis.* The raw *in situ* total scattering data were integrated using the software *Dioptas*,[89] while total scattering pair distribution functions (PDFs) were obtained from the integrated data using *PDFgetX3*.[90] A background scattering pattern obtained from a capillary containing deionized water at the corresponding conditions (pressure and temperature) was subtracted from the total scattering patterns prior to Fourier transformation. The $Q$-range employed in the Fourier transform was limited to 0.9–18 Å$^{-1}$ due to poor counting statistics at higher scattering vectors. Selected PDFs were modeled using the *PDFgui* software.[91] The given $MFe_2O_4$ ($M$ = Mn, Co, Ni, Zn) structures were described in space group $Fd\bar{3}m$ and refined using the $r$-range 1–20 Å. Scale factor, crystallite size (spherical particle diameter), unit cell, and atomic displacement parameters were refined.

*In Situ PXRD Data Treatment and Sequential Refinement.* The raw *in situ* PXRD data frames were integrated using the *Fit2D* software,[92] and sequential Rietveld refinement was carried out using the *Fullprof Suite* software package.[93] The given $MFe_2O_4$ ($M$ = Mn, Co, Ni, Zn) structures were described in space group $Fd\bar{3}m$. In specific data sets, a secondary hematite ($\alpha$-$Fe_2O_3$, space group $R\bar{3}c$) phase was observed and implemented in the refinement. The refinements of each series were done sequentially backward in time, starting from the final frame. The peak profiles were modeled using the Thompson–Cox–Hastings formulation of the pseudo-Voigt function.[94] The instrumental contribution to the peak profiles were determined by Rietveld refinement of data from a NIST $LaB_6$ 660B line profile standard and corrected for in the refinements. The atomic structures (atomic positions, site occupation fractions, displacement factors) were fixed based on our recent high-resolution neutron powder diffraction study.[68] The zero shift was refined for the last frame and held fixed in the sequential refinement, while the background (Chebyshev polynomial), scale factors, lattice parameters and one peak shape parameter related to isotropic crystallite size broadening were refined throughout the entire time-resolved data set.

## ASSOCIATED CONTENT

### ⓈSupporting Information

The Supporting Information is available free of charge at https://pubs.acs.org/doi/10.1021/acsnano.3c08772.

> Contour plots of *in situ* synchrotron powder X-ray diffraction data collected during the hydrothermal synthesis of $MnFe_2O_4$, $CoFe_2O_4$, $NiFe_2O_4$, and $ZnFe_2O_4$ nanocrystallites at various reaction temperatures; Plots of lattice parameters as a function of reaction time extracted by Rietveld refinement from the *in situ* PXRD data; Discussion of the effect of size distributions on the obtained volume averaged mean crystallite dimensions from the peak profile analysis (PDF)

## AUTHOR INFORMATION

### Corresponding Author

**Henrik L. Andersen** – *Instituto de Ciencia de Materiales de Madrid (ICMM), CSIC, Madrid 28049, Spain; Facultad de Ciencias Físicas, Universidad Complutense de Madrid, Madrid 28040, Spain;* orcid.org/0000-0003-1847-8427; Email: henrik.andersen@csic.es

### Authors

**Cecilia Granados-Miralles** – *Instituto de Cerámica y Vidrio (ICV), CSIC, Madrid 28049, Spain;* orcid.org/0000-0002-3679-387X

**Kirsten M. Ø. Jensen** – *Department of Chemistry and Nanoscience Center, University of Copenhagen, København Ø 2100, Denmark;* orcid.org/0000-0003-0291-217X

**Matilde Saura-Múzquiz** – *Facultad de Ciencias Físicas, Universidad Complutense de Madrid, Madrid 28040, Spain;* orcid.org/0000-0002-3572-7264

**Mogens Christensen** – *Department of Chemistry and Interdisciplinary Nanoscience Center, Aarhus University, Aarhus C 8000, Denmark;* orcid.org/0000-0001-6805-1232

Complete contact information is available at:
https://pubs.acs.org/10.1021/acsnano.3c08772

### Notes

The authors declare no competing financial interest.

## ACKNOWLEDGMENTS

The authors are grateful for the obtained beamtimes at beamline I711, MAXlab synchrotron radiation source, Lund University, Sweden; and beamline P02.1, PETRA III, DESY, Hamburg, Germany (ref. I-20150335 EC). Hanns-Peter Liermann, Jozef Bednarcik, Ann-Christin Dippel, Carsten Gundlach, Dörthe Haase, Diana Thomas, and Francisco Martinez are thanked for their support during the respective beamtimes. This work was financially supported by the Danish National Research Foundation (Center for Materials Crystallography, DNRF93), Innovation Fund Denmark (Green Chemistry for Advanced Materials, GCAM-4107-00008B), Independent Research Fund Denmark (Small and Smart Magnet Design) and the Danish Center for Synchrotron and Neutron Science (DanScatt), The Spanish Ministry of Universities (Ministerio de Universidades) and the European Union—NextGenerationEU through a Maria Zambrano—attraction of international talent fellowship grant, and Comunidad de Madrid, Spain, through an "Atracción de Talento Investigador" fellowship (2020-T2/IND-20581). This project has received funding from the European Union's Horizon Europe research and innovation programme under project No. 101063369 (OXYPOW). C.G.-M. acknowledges financial support from grant RYC2021-031181-I funded by MCIN/AEI/10.13039/501100011033 and by the "European Union NextGenerationEU/PRTR". C.G.-M. and M.S.-M. also acknowledge support from the Spanish Ministry of Science and Innovation through Grant TED2021-130957B-C51 and TED2021-130957B-C52 (NANOBOND) funded by MCIN/AEI/10.13039/501100011033 and by the "European Union NextGenerationEU/PRTR". Affiliation with the Center for Integrated Materials Research (iMAT) at Aarhus University is gratefully acknowledged.